\documentclass[journal=jctcce,manuscript=article,layout=twocolumn]{achemso}
\setkeys{acs}{articletitle = true}

\usepackage{bm,amsmath}
\usepackage{graphicx}
\usepackage{color}
\usepackage{caption}
\usepackage{subcaption}
\usepackage{booktabs}
\usepackage{footnote}
\usepackage{multirow}
\usepackage[dvipsnames]{xcolor}
\usepackage{algorithmic}
\usepackage[linesnumbered]{algorithm2e}
\usepackage{float}
\usepackage{braket}
\usepackage{cuted}
\usepackage{arydshln}
\usepackage{mathtools}
\usepackage{tabls}
{}
\usepackage[fontsize=10pt]{scrextend}
\usepackage[version=3]{mhchem}
\usepackage{hyperref}
\usepackage{cleveref}
\usepackage{tikz}
\usepackage{multirow}
\usepackage[shortlabels]{enumitem}

\graphicspath{{./figures/}}

\let\oldnl\nl% Store \nl in \oldnl
\newcommand{\nonl}{\renewcommand{\nl}{\let\nl\oldnl}}

\SectionNumbersOn{}

\author{Elise Palethorpe}
\affiliation{School of Computing, Australian National University, Canberra, ACT 2601, Australia}
\author{Ryan Stocks}
\affiliation{School of Computing, Australian National University, Canberra, ACT 2601, Australia}
\author{Giuseppe M. J. Barca}
\email{giuseppe.barca@unimelb.edu.au}
\affiliation{School of Computing and Information Systems, Melbourne University, Melbourne, VIC 3052, Australia}

\title{Advanced Techniques for High-Performance Fock Matrix Construction on GPU Clusters}
\begin{document}

\begin{abstract}
This Article presents two optimized multi-GPU algorithms for Fock matrix construction, 
building on the work of \citeauthor{ufimtsev_quantum_2009}~\cite{ufimtsev_quantum_2009} and \citeauthor{barca_faster_2021}~\cite{barca_faster_2021} 
The novel algorithms, opt-UM and opt-Brc, introduce significant enhancements, 
including improved integral screening, exploitation of sparsity and symmetry, 
a linear scaling exchange matrix assembly algorithm, and extended capabilities 
for Hartree-Fock caculations up to $f$-type angular momentum functions. 
Opt-Brc excels for smaller systems and for highly contracted triple-$\zeta$ basis sets, 
while opt-UM is advantageous for large molecular systems.  

Performance benchmarks on NVIDIA A100 GPUs show that our algorithms in the EXtreme-scale Electronic Structure System (\texttt{EXESS}), when combined,  outperform all current GPU and CPU Fock build implementations in \texttt{TeraChem}, \texttt{QUICK}, \texttt{GPU4PySCF}, \texttt{LibIntX}, \texttt{ORCA}, 
and \texttt{Q-Chem}. 
The implementations were benchmarked on linear and globular systems and average speed ups across three double-$\zeta$ basis sets of 1.5$\times$, 5.2$\times$, and 8.5$\times$ 
were observed compared to \texttt{TeraChem}, \texttt{GPU4PySCF}, and \texttt{QUICK} respectively. 

Strong scaling analysis reveals over 91\% parallel efficiency on four GPUs for opt-Brc, 
making it typically faster for multi-GPU execution. 
Single-compute-node comparisons with CPU-based software like \texttt{ORCA} and \texttt{Q-Chem} show speedups of up to 42$\times$ 
and 31$\times$, respectively, enhancing power efficiency by up to 18$\times$.

\end{abstract}

%=====================
\section{Introduction}
%=====================

The evolution of quantum chemical calculations relies heavily on the simultaneous advancement of both underlying algorithms and the computer systems they employ. The last two decades have witnessed a fundamental shift in computing, driven by the end of Dennard scaling and the gradual slowdown of Moore’s law. These changes mark the decline of the era dominated by general-purpose processors. To surpass the performance limitations of general-purpose systems, cutting-edge supercomputing platforms have adopted heterogeneous architectures. In these systems, CPUs primarily handle flow control, while specialized accelerators take on the bulk of computational tasks.

Among the various accelerators, graphics processing units (GPUs) have emerged as the most widespread and successful. However, leveraging GPUs to achieve substantial performance improvements over CPU-only computations is far from straightforward. The unique architecture of GPUs necessitates significant investment in new programming models and often requires a complete redesign of algorithms to optimize the software for the hardware.\cite{gordon2020novel} This shift represents a pivotal moment in the evolution of computational chemistry, where the integration of novel hardware and algorithmic innovation is essential for continued progress.

Within the hierarchy of quantum chemistry approaches, the Hartree-Fock (HF) method is foundational.\cite{szabo_modern_1996} It provides the theoretical and algorithmic framework for Kohn-Sham Density Functional Theory (KS-DFT)\cite{kohn_self-consistent_1965} and more accurate post-Hartree-Fock wave-function methods, such as M\"{o}ller-Plesset perturbation theory\cite{moller_note_1934} and coupled-cluster theories.\cite{szabo_modern_1996} This positions HF as a strategic target for both hardware advancements and algorithmic refinements that can drive the field forward.

The primary computational bottleneck of HF is the construction of the Fock matrix, which requires the evaluation of two-electron repulsion integrals (ERIs) and their subsequent digestion with the electron density matrix. 
The number of ERIs scales as $\mathcal{O}(N^4)$ with system size, presenting a significant computational challenge. 
Extensive research has focused on mitigating the steep computational cost of Fock matrix construction. 
This includes the development of integral screening techniques to reduce its formal complexity to $\mathcal{O}(N^2)$ 
\cite{whitten_coulombic_1973, haser_improvements_1989, sabin_molecular_1994, barca_two-electron_2016, barca_three-_2017, thompson_distance-including_2017, black_avoiding_2023}, 
the use of the continuous fast multipole method (CFMM) to obtain the Coulomb matrix ($\bm{J}$) in $\mathcal{O}(N\log{N})$ time, 
and approaches like LinK \cite{ochsenfeld_linear_1998}, the chain-of-spheres exchange (COSX) method \cite{neese_efficient_2009}, 
and the seminumerical exchange (sn-K) method\cite{laqua_highly_2020} that integrate optimal screening of the exchange repulsion interactions 
to achieve linear time complexity for the $\bm{K}$ matrix construction.

Despite their potential, these approaches are limited in significantly reducing overall computational time and enhancing scalability unless they are combined with novel algorithms designed to leverage the computational capacity of modern parallel computing architectures. This necessity has driven substantial research over the past two decades to develop new HF algorithms that can efficiently utilize the power of GPUs.\cite{yasuda_two-electron_2008,ufimtsev_quantum_2008, ufimtsev_quantum_2009-1,luehr_dynamic_2011,asadchev_new_2012,miao_acceleration_2013,yasuda_efficient_2014, rak_brush_2015, kussmann_hybrid_2017,kussmann_employing_2017,kalinowski_arbitrary_2017,barca_high-performance_2020,laqua_highly_2020, manathunga_harnessing_2021, seritan_terachem_2021, tian_optimizing_2021, johnson_multinode_2022,seidl2022q,manathunga_quantum_2023,qi_hybrid_2023,williams-young_distributed_2023,asadchev_high-performance_2023,galvez_vallejo_high-performance_2022,hicks_massively_2024,wu_python-based_2024,stocks2024multi}

In 2008, Yasuda implemented the first GPU algorithm for evaluating two-electron integrals involving $s$ and $p$-type Gaussian functions  and forming the Coulomb matrix $\bm{J}$ \cite{yasuda_two-electron_2008}. Using the 
McMurchie-Davidson (MMD)\cite{mcmurchie1978one} recursive scheme, each GPU thread computed a two-center primitive integral class $[p|q]$ and contracted its constituent  ERIs with density matrix elements until the Coulomb matrix was completed, without exploiting ERI symmetry.

Following Yasuda's seminal work, Ufimtsev and Martinez developed a full Fock build implementation for GPUs \cite{ufimtsev_quantum_2008, ufimtsev_quantum_2009-1}, also based on the MMD recursive scheme. This implementation, herein referred to 
as UM09, significantly improved workload balance and overall performance by organizing ERI data in blocks of uniform angular momentum and employing two algorithms: a $\bm{J}$-engine-based\cite{white1996aj} Coulomb matrix algorithm and an 
exchange matrix $\bm{K}$ algorithm. Both algorithms partially sacrificed symmetry exploitation to prevent inter-GPU-block synchronization and memory bank conflicts. 

In 2012, Asadchev and Gordon introduced a Fock build algorithm utilizing a Rys quadrature ERI algorithm,\cite{dupuis_evaluation_1976,rys_computation_1983} enabling evaluations to extend to  $g$-type Gaussians \cite{asadchev_new_2012}. They utilized eight-fold permutational symmetry and assigned different GPU warps to update different 
Fock matrix blocks. Although this reduced re-computation, significant thread synchronization was necessary to avoid race conditions using mutual exclusion objects.

In 2013, Miao and Merz proposed an approach where contracted integral classes were mapped to different GPU threads, with the main innovation being the use of the Head-Gordon-Pople (HGP) algorithm to minimize the computational cost 
associated with the evaluation of a contracted integral class \cite{miao_acceleration_2013}. This approach was implemented in the \texttt{QUICK} GPU program, and further extended to multi-GPU parallelism.\cite{manathunga_harnessing_2021, manathunga_quantum_2023,manathunga_quick-2403_2024}

In 2020, Barca \emph{et al.} introduced a novel algorithm that also mapped contracted integral classes to GPU threads,\cite{barca_high-performance_2020} fine-tuning this scheme and extending it to 
multi-GPU and multi-node parallelism in 2021.\cite{barca_faster_2021} This scheme, herein referred to as Brc21, was also based on HGP and incorporated algorithmic innovations such as efficient use of integral screening, 
full exploitation of integral symmetry, a novel ERI digestion scheme, and various GPU-specific optimizations that enabled it to outperform existing GPU and CPU 
implementations. The original Brc21 scheme was implemented in the LibCChem library of \texttt{GAMESS},\cite{barca2020recent,zahariev2023general} and then a distinct version was implemented in the EXtreme-scale Electronic Structure System (\texttt{EXESS})\cite{barca_enabling_2021,barca2022scaling,galvez_vallejo_toward_2023}. 

In the last two years, Asadchev and Valeev observed that the shift from compute-bound to memory-bound for integral classes involving $d$-type functions in Brc21\cite{barca_faster_2021} might allow a matrix-based formulation of the MMD algorithm to outperform it for high-angular momentum functions.\cite{asadchev_high-performance_2023} Recently they extended this MMD scheme to low-angular-momentum functions.\cite{asadchev_3-center_2024} However, the current implementations of Asadchev and Valeev's work in \texttt{LibIntX} provide only ERI evaluation and an initial $\bm{K}$ matrix assembly, and therefore do not yet allow for full performance verification. Furthermore, whether the MMD approach in \texttt{LibIntX} can outperform UM09 for $d$ and $f$ functions remains to be established due to the lack of a full Fock build algorithm.
In Section 6.1 of this paper, we will provide evidence that for up to $d$-functions, an optimized algorithm based on the UM09 scheme for the $\bm{K}$-matrix assembly designed by us significantly outperforms the $\bm{K}$-matrix assembly algorithm in \texttt{LibIntX}.

Recently, Li \emph{et al.} introduced a single GPU algorithm integrated into \texttt{GPU4PySCF},\cite{li_introducing_2024} the GPU-accelerated version of \texttt{PySCF}. This implementation is based on the Rys Quadrature ERI evaluation scheme and uses the Brc21 scheme for integral class creation and screening \cite{dupuis_evaluation_1976,rys_computation_1983}. 
While the use of Rys quadrature is rationalized as providing a lower GPU memory footprint, we demonstrate in this paper that it results in significantly lower performance compared to optimized versions of the Brc21 and UM09 schemes.

Thus, to date, each of these GPU implementations is based on a single algorithm tailored on a specific recursive scheme, either MMD, HGP, or Rys. Learning from established literature on ERI calculations on CPU,\cite{gill_molecular_1994} it is unlikely for a single algorithm to be optimal across different molecular systems, basis sets with different angular momenta, and different numbers of GPUs employed. 

To date, the UM09 and Brc21 implementations have shown the most promise in practical performance benchmarks. There are significant differences in how each of these schemes addresses the challenges of efficient ERI evaluation and digestion. For example, the Brc21 scheme fully exploits the eight-fold permutational symmetry of the ERIs, while the UM09 scheme elects to split the Coulomb and exchange computation for fine-grained density screening. Additional differences include the ERI evaluation recurrence relation, degree of parallelism, and shell pair batching strategy. In general, however, the UM09 scheme optimizes for efficient screening while the Brc21 scheme minimizes computation per ERI.

In this Article, we provide improved multi-GPU algorithms for both the Brc21 and UM09 schemes, named opt-Brc and opt-UM. This enables us to provide evidence that the fastest algorithm requires executing one or the other schemes, but neither of the two dominates in performance across the whole range of basis sets, system sizes, and the number of GPUs utilized. 

We also show that, when combined, the opt-Brc and opt-UM schemes significantly outperform all current GPU and CPU Fock build implementations in \texttt{Terachem}, \texttt{QUICK}, \texttt{GPU4PySCF}, \texttt{LibIntX}, \texttt{ORCA} and \texttt{Q-Chem}.

Besides various algorithmic optimizations and novel implementations of the opt-Brc and opt-UM schemes, additional contributions of this work are:
\begin{itemize}
    \item Improvements to the efficiency and effectiveness of screening in the Brc21 Fock build.
    \item Extension of the UM09 Fock build scheme to exploit the exponential decay of the density matrix to improve screening and yield a linear scaling $\bm{K}$-matrix assembly algorithm. 
    \item Extension of the code capabilities up to $f$-type Gaussian basis functions to consider more diverse chemical systems.
\end{itemize}

All novel algorithms were implemented in the Extreme-scale Electronic Structure System (\texttt{EXESS}).\cite{galvez_vallejo_toward_2023}

The remainder of this Article is structured as follows. Section \ref{sec:background} details the notation and the primary algorithmic challenges of the HF method. Sections \ref{sec:barca-opt} and \ref{sec:um-scheme} detail how the Brc21 and UM09 schemes respectively address the algorithmic challenges and our novel optimisations to these schemes. Section \ref{sec:Code generators} briefly details our automatic code generation framework, followed, in Section \ref{sec:results}, by a performance analysis and extensive comparisons with GPU and CPU implementations in widely-used quantum chemistry packages.

\section{Theoretical Background}\label{sec:background}
%===============
\subsection{Notation}\label{sec:notation}
%===============
We largely follow the notation presented by \citeauthor{barca_faster_2021}\cite{barca_faster_2021}  A contracted Gaussian function (CGF), denoted by its basis set index $\alpha$,
\begin{equation}\label{eq:cgf}
    |\alpha) \equiv \phi_{\alpha}(\bm{r}) = \sum_{i=1}^{K} \varphi^{\alpha}_i(\bm{r})
\end{equation}
is defined as a sum of $K$ primitive Gaussian functions (PGFs)
\begin{multline} 
\label{eq:prim}
	|\alpha]_i \equiv \varphi^{\alpha}_i(\bm{r}) = \\ D_i (\bm{r}_x-A_x)^{a_x} (\bm{r}_y-A_y)^{a_y} (\bm{r}_z-A_z)^{a_z} e^{-\lambda_{i} | \bm{r}-\bm{A} |^2}
\end{multline}
where $\bm{a}=(a_x, a_y, a_z)$ in an angular momentum vector, $a = a_x + a_y + a_z$ is the total angular momentum, $\bm A \equiv (A_{x},A_{y},A_{z})$ is the Cartesian coordinate centre, $D_i$ is a contraction coefficient, and $\lambda_i$ the primitive Gaussian exponent. For the sake of brevity, we will often omit the primitive index $i$.

It is useful to overload the notation for these basis functions using their angular momentum vectors as follows
\begin{align}
        |\bm{a}) &\equiv |\alpha), \\
        |\bm{a}] &\equiv |\alpha].
\end{align}

For computational convenience Gaussian functions are grouped into shells. A primitive shell will be indicated using the non-bold notation $|a]$, and is defined as the set of primitives $|\bm{a}]$ sharing the same total angular momentum $a$, primitive exponent $\lambda_i$ and centre $\bm{A}$. Similarly, a contracted shell $|a)$ is a set of contracted Gaussian functions $|\bm{a})$ sharing the same set of primitive exponents $\{\lambda_{i}\}$, the same centre $\bm{A}$ and total angular momentum $a$. There are $\frac{(a + 1)(a + 2)}{2}$ Cartesian contracted basis functions in a shell with total angular momentum $a$. For example, a $|p)\equiv|1)$ shell represents a set of three $p$-type CGFs, namely $\{|\bm{a}_{1})=(1,0,0),|\bm{a}_{2}) =(0,1,0),|\bm{a}_{3})=(0,0,1)\}$, which correspond to the familiar $\bm{p}_x$, $\bm{p}_y$, and $\bm{p}_z$ functions, respectively.  

A CGF pair 
\begin{equation}
 	|\alpha\beta) \equiv |\bm{a}\bm{b}) = \phi_{\alpha}(\bm{r}) \phi_{\beta}(\bm{r})
 \end{equation}
 is a sum of PGF pairs 
 \begin{equation}
  	|\alpha\beta] \equiv |\bm{a}\bm{b}] = \varphi_{\alpha}(\bm{r}) \varphi_{\beta}(\bm{r}).
 \end{equation}

Analogously, a contracted shell pair is the set of contracted basis function pairs obtained by the tensor product $|ab) = |a) \otimes |b)$. Primitive shell pairs are defined similarly, $|ab] = |a] \otimes |b]$.
Primitive and contracted shell pairs are further coupled into
shell quartets $|abcd] = |ab] \otimes |cd]$ and $|abcd) = |ab) \otimes |cd)$.

Furthermore, we will adopt the following curly braces notation $|ab\}$ to indicate a batch of contracted shell pairs $|ab)$. 

Central to the Hartree-Fock method is the evaluation of the four-centre electron repulsion integrals (ERIs)
\begin{equation}
    (\alpha \beta \vert \gamma \delta) \equiv (\bm{ab} \vert \bm{cd}) 
 = \sum_{i}^{K_{A}}\sum_{j}^{K_{B}}\sum_{k}^{K_{C}}\sum_{l}^{K_{D}} [\bm{a}\bm{b}|\bm{c}\bm{d}]_{ijkl}
\end{equation}
where
\begin{multline}
	[\alpha\beta|\gamma\delta]\equiv [\bm{ab} | \bm{cd}]_{ijkl} =\\ \iint \varphi^{\alpha}_i(\bm{r}_1) \varphi^{\beta}_j(\bm{r}_1)\, \frac{1}{|\bm{r}_{1}-\bm{r}_{2}|}\, \varphi^{\gamma}_k(\bm{r}_2) \varphi^{\delta}_l(\bm{r}_2) d \bm{r}_1 d \bm{r}_2.
\end{multline}

We use the notation $(ab|cd)$ to indicate the set of ERIs over all basis functions quartets arising from the tensor product of two shell pairs $|ab) \otimes |cd)$. For example, $(11|11)\equiv (pp|pp)$ is the set of 81 contracted integrals $\{(\bm{p}_x\bm{p}_x|\bm{p}_x \bm{p}_x), (\bm{p}_x\bm{p}_x|\bm{p}_x\bm{p}_y), \ldots, (\bm{p}_z\bm{p}_z|\bm{p}_z\bm{p}_z) \}$.

ERIs are generally evaluated using recurrence relations starting from the following fundamental integral
\begin{multline}
    [\bm{00}\vert\bm{00}] =\\ U_{P}U_{Q}\iint e^{-\zeta\, |\bm{r}_{1}-\bm{P}|^{2}}\, \frac{1}{\bm{r}_{1}-\bm{r}_{2}}\, e^{-\eta\, |\bm{r}_{2}-\bm{Q}|^{2}}\, d\bm{r}_1 d\bm{r}_2 
\end{multline}

where
\begin{gather}
    \label{eq:zeta}
    \zeta = \lambda_{i}+\lambda_{j} \hspace{1cm} \eta = \lambda_{k}+\lambda_{l}\\
    \bm{P} = \frac{\lambda_{i} \bm{A} + \lambda_{j}\bm{B} }{\zeta} \hspace{1cm}
    \bm{Q} = \frac{\lambda_{k}\bm{C}  + \lambda_{l}\bm{D} }{\eta}\\
    U_{P} = n_{a}n_{b}\, D_i D_j\, \left(\frac{\pi}{\zeta}\right)^{3/2}\, e^{- \frac{\lambda_{i}\lambda_{j}}{\zeta} |\bm{A}\bm{B}|^{2}} \\
    U_{Q} = n_{c}n_{d}\, D_k D_l\, \left(\frac{\pi}{\eta}\right)^{3/2}\, e^{- \frac{\lambda_{k}\lambda_{l}}{\eta} |\bm{C}\bm{D}|^{2}}.
\end{gather}
Here, $n_{a}$ is a normalization factor and  $|\bm{A}\bm{B}|=|\bm{A}-\bm{B}|$.

The following auxiliary integrals are also required for the recursive evaluation of the ERIs

\begin{equation}
    [\bm{00}|\bm{00}]^{(m)} = U_PU_Q\theta^m F_m(T)
\end{equation}
where
\begin{align} \label{eq:boys-fn-def}
    F_m(T)&=\int_0^1 t^{2m} \exp(-t^2 T) dt
\end{align}
is the generalized Boys function, $\theta = \sqrt{\frac{\zeta\eta}{\zeta + \eta}}$ and $T=\theta^{2}|\bm{PQ}|^2$. 

When discussing recurrence relations, which are needed to evaluate ERIs that include non-\emph{s} functions, we use a compact vector notation, wherein $\bm{a}^\pm$ represents the three angular momentum vectors formed by incrementing (+) or decrementing (-) the $x$, $y$, or $z$ components of $\bm{a}$. The remaining vectors in the equation should also be taken in that component (\emph{e.g.}, if decrementing $\bm{a}$ in the $x$ direction, $\bm{PA}$ should be read as $PA_x$).
For example, using this notation, Boys’ famous formula for the derivatives of the primitive Gaussian $[\bm{a}|$ with respect to the components of its centre $\bm{A}$ appears as follows

\begin{equation}
\nabla_{\bm{A}} [\bm{a}| = 2 \lambda [\bm{a}^{+}| - \bm{a}^{-} [\bm{a}^{-}|.
\end{equation}

For efficient ERI evaluation, integrals are grouped by their angular momentum and potentially contraction degree. We will refer to this grouping as a class of integrals or a sub-class when integrals are also grouped by contraction degree.

%===============
\subsection{Hartree-Fock Energy Formulation}\label{sec:HF-notation}
%===============

In the Hartree-Fock (HF) method, the minimum energy Slater determinant is found by solving the following generalized eigenvalue problem
\begin{equation}
    \boldsymbol{FC = SC\epsilon},
\end{equation}
where $\boldsymbol{F}$ is the Fock matrix in the atomic orbital (AO) basis, $\boldsymbol{C}$ are the molecular orbital (MO) coefficients, $\boldsymbol{S}$ is the AO overlap matrix and $\boldsymbol{\epsilon}$ is a diagonal matrix representing the MO energies. 

The Fock matrix is itself a function of the MO coefficients so the equations are solved iteratively through the Self Consistent Field (SCF) procedure until convergence. Note that here we will present the formulation using the restricted closed shell approach. Adaptation to unrestricted HF is straightforward, but requires maintaining two copies of many of the involved  matrices.

The focus of this Article is the primary bottleneck of an HF calculation, which is the construction of the Fock matrix 
\begin{equation}
\label{eq:fock_build}
    F_{\mu \nu} = h_{\mu\nu} + J_{\mu\nu} - \frac{1}{2}K_{\mu\nu},
\end{equation}
where $h_{\mu\nu}$ is the atomic orbital representation of the core Hamiltonian, $D_{\mu\nu}$ is the doubly occupied density matrix
\begin{equation}
    D_{\mu\nu} = 2\sum_i C_{i\mu}C_{i\nu},
\end{equation}
and 
\begin{align}
        J_{\mu \nu} &= \sum_{\lambda \sigma} (\mu\nu|\lambda\sigma) D_{\lambda\sigma}, \\
        K_{\mu \nu} &= \sum_{\lambda \sigma} (\mu\lambda|\nu\sigma) D_{\lambda\sigma}, 
\end{align}
are the elements of the Coulomb ($\bm{J}$) and exchange repulsion ($\bm{K}$) matrices, respectively.

The computation of $h_{\mu\nu}$ is relatively cheap and performed only once during the SCF so the calculation of the four-center ERIs, and their ensuing combination with the density matrix to form Fock matrix elements, the ERIs digestion, constitute the primary bottleneck of the Fock build. The four center ERIs require too much memory to be stored, and for computational efficiency are re-computed at every iteration of the SCF procedure.

In the absence of any further algorithmic refinement, the computation of both $\bm{J}$ and $\bm{K}$, and therefore of $\bm{F}$ scales as $\mathcal{O}(N^4)$. To mitigate the steep computational cost of the Fock matrix construction, integral screening techniques have been developed to reduce its formal complexity to $\mathcal{O}(N^2)$.\cite{whitten_coulombic_1973,haser_improvements_1989,gill_molecular_1994,barca_two-electron_2016,barca_three-_2017,thompson_distance-including_2017, black_avoiding_2023}. Arguably the most commonly adopted of these ERI screening methods is exploiting the following Cauchy Schwarz inequalities
\begin{equation} \label{eq:cauchy-schwarz}
    |(\bm{ab}\vert \bm{cd})| \le G_{ab}G_{cd} 
\end{equation}
\begin{equation} 
\label{eq:cauchy-schwarz2}
    |[\bm{ab}\vert \bm{cd}]| \le G_{[ab]}G_{[cd]} 
\end{equation}
where $G_{ab} = \max_{|\bm{ab})\in|ab)} \sqrt{ (\bm{ab}|\bm{ab})}$, and $G_{[ab]} = \max_{|\bm{ab}]\in|ab]} \sqrt{ [\bm{ab}|\bm{ab}]}$ . Efficiently utilising Eqs. \eqref{eq:cauchy-schwarz} and \eqref{eq:cauchy-schwarz2}  specifically on GPU hardware for ERI screening purposes is a key objective of this Article.

\subsection{Algorithmic Challenges}
Designing an efficient, distributed-memory and GPU accelerated implementation of the Fock build poses several computational challenges.

\begin{itemize}
\item \textbf{Efficient Evaluation of ERIs}. The most efficient algorithms for the evaluation of ERIs compute all integrals arising from a shell quartet simultaneously. There is significant overlap in the recursive intermediates required for evaluation of integrals arising from the same shell quartet. Therefore, to reduce FLOPs, the integrals from each shell quartet are computed solving a tree search problem to minimize the number of recursive intermediates stored for reuse. However, this results in very high register usage which can reduce the achieved warp occupancy on the GPU or lead to local memory spills resulting in poor performance. 
\item \textbf{Screening}. For a high-performance Fock build, it is essential to avoid the computation of numerically insignificant integrals, for example via the Cauchy Schwartz inequality (Eq.~\eqref{eq:cauchy-schwarz}). However, this conditional screening has the potential to lead to significant warp divergence when implemented on GPU. 
\item \textbf{Symmetry}. The third challenge arises from fully exploiting the 8-way permutational symmetry of the ERIs 
\begin{multline}
(\mu\nu|\lambda\sigma)=(\nu\mu|\lambda\sigma)=(\mu\nu|\sigma\lambda)=(\nu\mu|\sigma\lambda)\\=(\lambda\sigma|\mu\nu)=(\sigma\lambda|\mu\nu)=(\lambda\sigma|\nu\mu)=(\sigma\lambda|\nu\mu).
\end{multline}
Harnessing this symmetry can lead to significant computational savings. However, if only symmetry unique integrals are computed, each integral must contribute to the formation of multiple matrix elements. This can give rise to race conditions adding synchronisation costs. Additionally, full symmetry utilisation forces screening based on the density matrix to be less fine grained.
\item \textbf{Load Balancing}. The evaluation of integrals is conducted at the granularity of a class rather than individually. Significantly more integrals arise from shell quartets with high angular momentum than those with low angular momentum. Furthermore, within the same angular momentum category, sub-classes involving more highly contracted CGFs are much more computationally expensive than those with a lower degree of contraction. Finally, the number of integrals per class varies greatly with angular momentum and contraction degree. This computational heterogeneity poses a load balancing challenge when distributing the work across multiple GPUs.
\end{itemize}

These algorithmic challenges have been tackled in the literature by significantly different approaches, with the highest-performance being the UM09 scheme and the Brc21 scheme. 
We will now discuss these two algorithms, and how to improve upon them to achieve better performance. 

%===============
\section{Brc21 scheme optimisations}\label{sec:barca-opt}
%===============
The Brc21 scheme is a GPU-tailored algorithm based on the Head-Gordon-Pople (HGP) scheme for ERI evaluation\cite{barca_faster_2021}. In this scheme, the contraction of primitive integrals is performed midway. Initially, the angular momentum is built on two centers using primitive recurrence relations, followed by contraction. The final calculation of the target integrals is completed using contracted integral recurrence relations. This approach has been shown to minimize FLOP counts for various contracted integral classes compared to late-contraction schemes, which perform the contraction of the primitive integrals only at the very end, after building angular momentum on all four Gaussian centers using primitive recurrence relations~\cite{gill_molecular_1994}.

A one-thread-per-contracted-ERI approach is used to maximize parallelism. Partial pre-digestion in thread-private memory and less fine-grained density screening are employed to exploit the 8-fold ERI symmetry while keeping thread synchronization overhead low.

\subsection{HGP Recursive Scheme}
Let us start by presenting the HGP refinement of the Obara-Saika recurrence scheme~\cite{headgordon_method_1988}.

The target contracted integrals $(\bm{ab}|\bm{cd}) \in (ab\vert cd)$ are calculated starting from the primitive fundamental integrals $[\bm{00}|\bm{00}]^{m}$, with $m \in \{0,\ldots,a+b+c+d\}$, and then applying the following vertical recurrence relations (VRRs)
\begin{align}
\label{eq:vrr1}
        [\bm{e}^{+}\bm{0}|\bm{f}\bm{0}]^{(m)} &=  \bm{PA}\,  [\bm{e}\bm{0}|\bm{f}\bm{0}]^{(m)} 
         -  \bm{PQ}\frac{\eta}{\zeta+\eta} [\bm{e}\bm{0}|\bm{f}\bm{0}]^{(m+1)} \nonumber \\
        & +  \frac{\bm{e}}{2\zeta} \left( [\bm{e}^{-}\bm{0}|\bm{f}\bm{0}]^{(m)} 
         -  \frac{\eta}{\zeta+\eta} 
        [\bm{e}^{-}\bm{0}|\bm{f}\bm{0}]^{(m+1)} \right) 
        \nonumber \\
        & + \frac{\bm{f}}{2(\zeta+\eta)}\,  [\bm{e}\bm{0}|\bm{f}^{-}\bm{0}]^{(m+1)} \hspace{3.1cm} 
\end{align}

\begin{align}
\label{eq:vrr2}
        [\bm{e}\bm{0}|\bm{f}^{+}\bm{0}]^{(m)} &=  \bm{QC}\,  [\bm{e}\bm{0}|\bm{f}\bm{0}]^{(m)} 
        -  \bm{PQ} \frac{\zeta}{\zeta+\eta} [\bm{e}\bm{0}|\bm{f}\bm{0}]^{(m+1)} \nonumber \\
        & + \frac{\bm{f}}{2\eta}  \left( [\bm{e}\bm{0}|\bm{f}^{-}\bm{0}]^{(m)} 
        -   \frac{\zeta}{\zeta+\eta} 
        [\bm{e}\bm{0}|\bm{f}^{-}\bm{0}]^{(m+1)} \right) 
        \nonumber \\
        & + \frac{\bm{e}}{2(\zeta+\eta)}\, [\bm{e^{-}}\bm{0}|\bm{f}\bm{0}]^{(m+1)} \hspace{3.1cm}
\end{align}

to obtain suitable intermediate primitive integrals $[\bm{e0}\vert\bm{f0}]$, which are contracted on the fly

\begin{equation}
     (\bm{e0}\vert \bm{f0}) =  \sum_{i}^{K_{A}} \sum_{j}^{K_{B}} \sum_{k}^{K_{C}} \sum_{l}^{K_{D}} [\bm{e0}\vert\bm{f0}]_{ijkl}
\end{equation}

and finally applying horizontal recurrence relations (HRRs)
\begin{subequations}
\label{eq:hrr}
\begin{align}
    (\bm{e}\bm{0}|\bm{c}\bm{d}^{+}) &= (\bm{e}\bm{0}|\bm{c}^{+}\bm{d}) + \bm{CD}\, (\bm{e}\bm{0}|\bm{c}\bm{d}) \label{eq:hrr1} \\
    (\bm{a}\bm{b}^{+}|\bm{c}\bm{d}) &= (\bm{a}^{+}\bm{b}|\bm{c}\bm{d}) + \bm{AB}\, (\bm{a}\bm{b}|\bm{c}\bm{d}). \label{eq:hrr2}
\end{align}
\end{subequations}

A key feature of the HGP scheme is that the HRRs in Eq.~\eqref{eq:hrr} do not depend on primitive-level data. Therefore, the HRRs can be used directly on contracted integrals---outside the contraction loops over $K_AK_BK_CK_D$---thereby greatly reducing the FLOP count if the target integrals involve sufficiently contracted and sufficiently high angular momentum CGFs. 

To minimize the computational cost associated with the algorithmic implementation of this recurrence scheme, it is necessary to solve a complex tree search problem for each angular momentum class. We solve these tree search problems using heuristic methods, implemented automatically via an in-house developed code generator, as detailed in Section \ref{sec:Code generators}.

\subsection{Optimised GPU Implementation}\label{sec:HGP screening}
\begin{figure*}
    \centering
    \includegraphics[width=\linewidth]{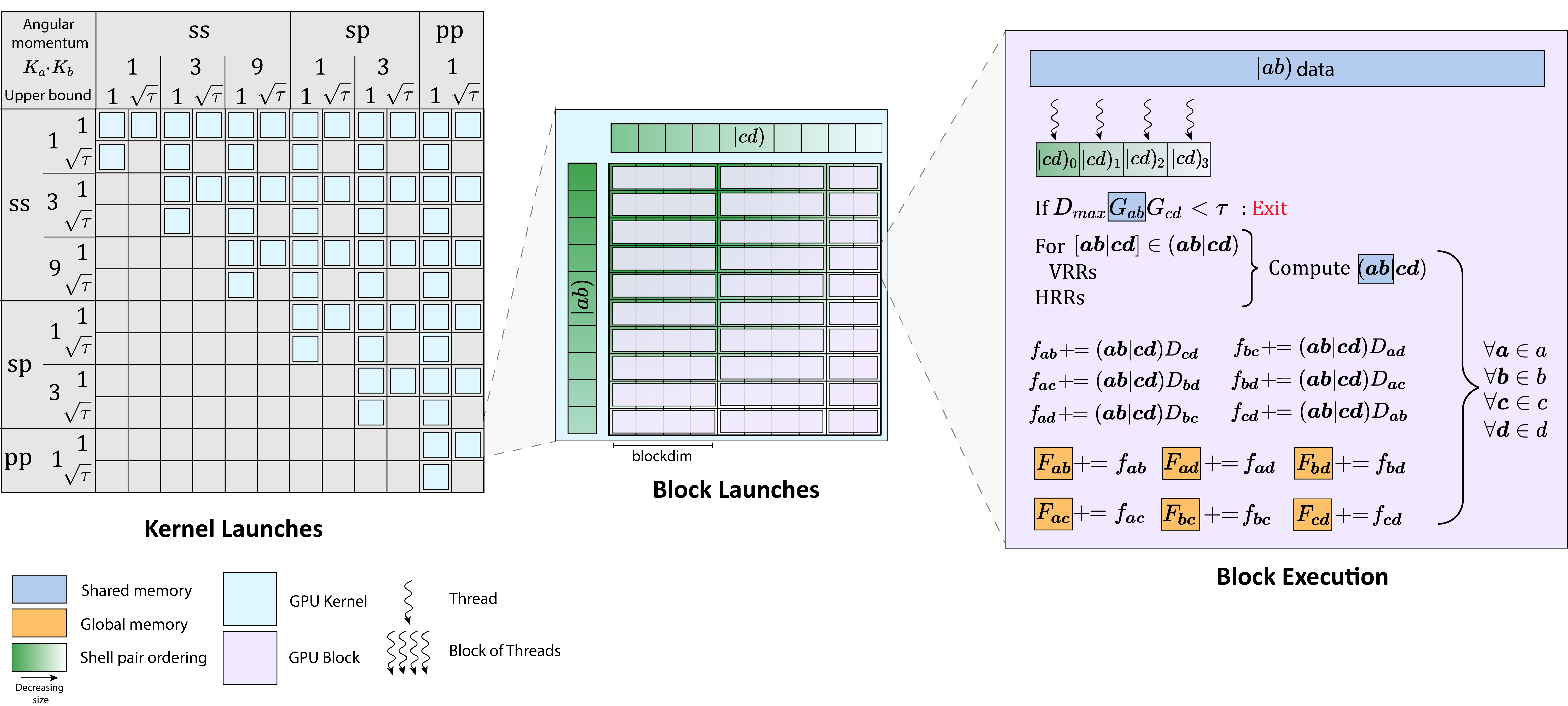}
    \caption{Algorithmic overview of the optimised Brc21 scheme.}
    \label{fig:HGP}
\end{figure*}
A scheme of our optimised GPU implementation of the Brc21 algorithm is shown in Fig. \ref{fig:HGP}.

To exploit the FLOP savings provided by the HGP approach, each thread is assigned to compute all contracted integrals arising from a single shell quartet. To maintain a good workload balance among GPU threads, all threads in a block are assigned shell quartets with the same angular momentum and contraction degree.

Significant shell pairs $|ab)$ are grouped into batches $|ab\}$ with uniform angular momentum and contraction degree. Specifically, to batch shell pairs, we select a list of batch magnitudes and assign each shell pair to the largest batch magnitude that is less than its $G_{ab}$ value. These groups are then sorted by the product of their Cauchy-Schwartz bound $G_{ab}$ and the maximum magnitude of the corresponding density matrix block to reduce warp divergence.

Constructing shell pair batches requires a careful balance between effective kernel screening, sufficient batches for dynamic load distribution, and adequate batch sizes to fully utilize the GPU.
Significant performance improvement was observed when the product of the shell-pair batch magnitude equates to the screening threshold value. This is achieved when the batch magnitudes are defined as:
\begin{equation}
    \tau ^ {\frac{i}{n}}
\end{equation}
where $\tau$ is a user-defined screening threshold, $n$ is chosen to optimize batch size, and $i$ is the magnitude of the $i^{\text{th}}$ batch. For all $i$ and $j$ such that $i+j \le n$, the batches will be screened. This pre-kernel-launch screening is advantageous as it eliminates kernel call overhead and any warp divergence associated with intra-kernel screening. 
All reported performance metrics in this paper were obtained with a target batch size of 2500. An example with $n=2$ is shown in the Kernel Launch component of Fig.~\ref{fig:HGP}, highlighting the significant number of kernel launches that can be avoided with this screening. 

Figure \ref{fig:HGP} also shows that the full 8-way permutational symmetry is exploited. This is exemplified in the Kernel Launch component of Fig. \ref{fig:HGP}, where $ps$-type shell-pair batches are not present and therefore neglected, and only kernels in the upper triangle of the shell quartet matrix are launched. For example, launching a $(sp|df)$ kernel avoids launching $(ps|df)$, $(sp|fd)$ and $(ps|fd)$ kernels by only considering shell pairs with monotonically increasing angular momentum, and avoids launching $(df|sp)$, $(df|ps)$, $(fd|sp)$ and $(df|ps)$ by restricting the launches to the upper triangle.

GPU thread blocks are launched such that all threads within a given block share the contracted shell pair $|ab)$. All data associated with $|ab)$ is stored in shared memory to achieve minimal latency and higher read bandwidth. The $|cd)$ shell pairs are assigned to blocks such that consecutive threads are allocated consecutive contracted shell pairs $|cd)$ within a batch $|cd\}$. 

Each GPU thread is responsible for computing all $(\bm{ab}|\bm{cd})\in (ab|cd)$ and updating the corresponding six Fock matrix blocks ($F_{ab}$, $F_{ac}$, $F_{ad}$, $F_{bc}$, $F_{bd}$, and $F_{cd}$) accounting for the 8 fold ERI symmetry as detailed in the Block Execution component of Fig. \ref{fig:HGP}.
In this strategy, multiple threads write to the same element of the Fock matrix. Thus, to avoid race conditions, these writes are performed atomically using the \texttt{atomicAdd} CUDA routine. 
To reduce the number of atomic operations, the ERIs are partially pre-digested with the density matrix into thread-private Fock matrix buffers, without the need of any synchronization. Without partial digestion, the number of atomic operations per thread would scale with the number of integrals arising from the shell quartet. Partial digestion reduces this significantly, only the two-index buffers require atomic operations. 

Screening before a kernel launch is advantageous, however, it cannot fully account for the magnitude of the density matrix. Due to significant sparsity within the density matrix, effective intra-kernel screening is necessary.
Each ERI is processed with six blocks of the density matrix, necessitating the computation of an ERI if any element within these six corresponding density blocks is significant. Before computing the ERI, we determine the maximum absolute element of the six density blocks $D_{max} = \max{\{|D_{ab}|,|D_{ac}|,|D_{ad}|,|D_{bc}|,|D_{bd}|,|D_{cd}|\}}$ and screen $(ab|cd)$ if $G_{ab}G_{cd}D_{max} < \tau$.
This has potential to lead to significant warp divergence. We therefore sort the batches of shell pairs by product of Cauchy Schwartz bound and maximum of the shell pair density block to increase likelihood that consecutive threads will return the same result from the screening conditional.
This ordering does not rigorously prevent warp divergence but our testing found a significant performance improvement over sorting only by the Cauchy Schwartz bound.

Screening also occurs at the primitive level with the primitive Cauchy Schwartz bound. When constructing contracted shell pairs we order the primitive pairs by their Cauchy Schwartz bound and delete insignificant primitive pairs to further reduce warp divergence and increase screening efficiency.

%===============
\section{UM09 scheme optimisations}\label{sec:um-scheme}
%===============
The UM09 scheme~\cite{ufimtsev_quantum_2009} enables an extremely efficient integral screening. This is achieved through an intelligent usage of McMurchie Davidson (MMD)~\cite{mcmurchie1978one} recurrence relations for computation of ERIs. Unlike the HGP scheme, the MMD scheme is a late contraction scheme. It is therefore preferable to use a one-thread-per-primitive approach as work between primitives can be efficiently parallelised, with screening performed at the primitive ERI level.

For fine density screening, complete symmetry utilization is sacrificed to enable screening based on one block of the density matrix rather than six.
This is achieved by separately computing the Coulomb ($\bm{J}$) and exchange contributions ($\bm{K}$) to the Fock matrix.  The Coulomb contribution can still exploit shell pair symmetry ($|ab] \leftrightarrow |ba]$ and $|cd] \leftrightarrow |dc]$), while the exchange contribution can exploit bra-ket symmetry ($[ab|cd] \leftrightarrow [cd|ab]$) with suitable symmetry post-processing.

Considering the exchange term separately is particularly beneficial, as the density elements involved in its calculation span the bra and the ket of the corresponding ERIs. It is well established that the magnitude of the AO density matrix elements decays exponentially with the distance between basis functions, a phenomenon known as the nearsightedness of the density matrix~\cite{kohn_analytic_1959, he_exponential_2001, maslen_locality_1998, ismail-beigi_locality_1999, barca_q-mp2-os_2020, goedecker_decay_1998}. Thus, for a given bra shell pair, only a constant number of ket shell pairs contribute significantly to the exchange matrix. If the molecular system is sufficiently large, exploiting this nearsightedness allows for a theoretically linear scaling formation of the exchange matrix.

Two performance concerns arise with this approach. First, in large systems, the proportion of Fock matrix contributions that can be screened increases dramatically. Thus, checking whether an integral can be screened (by loading the Cauchy-Schwarz bound and density block) can become a computational bottleneck. Second, if each thread computes a single primitive integral, atomic operations to increment the Fock matrix become the bottleneck of the algorithm, as there are significantly more writes to the global matrix compared to a contracted scheme.

To address these issues, we aim for each thread to compute multiple integrals where these integrals:
\begin{enumerate}[a)]
    \item Contribute to the same element of the Fock matrix.
    \item Are sorted in such a way that, once one integral is found to be insignificant, all remaining integrals are also insignificant, allowing computation to stop.
\end{enumerate}

The UM09 scheme can be optimized to meet both conditions effectively. We will begin by detailing the MMD scheme for ERI evaluation and then proceed to discuss the optimization of the evaluation of the $\bm{J}$ and 
$\bm{K}$ matrices on GPU.

\subsection{MMD Recursive Scheme}
In the McMurchie–Davidson method, expressions of the kind $(x-A_x )^{a_x} (x-B_x)^{b_x}$ arising from products of PGFs are expanded in a basis of Hermite polynomials, enabling the evaluation of the primitive ERIs as follows 
\begin{multline}
\label{eq:MD def1}
[\bm{a}\bm{b}\vert \bm{c}\bm{d}]=\\\sum_{p_x}^{a_x+b_x}\sum_{p_y}^{a_y+b_y}\sum_{p_z}^{a_z+b_z}\sum_{q_x}^{c_x+d_x}\sum_{q_y}^{c_y+d_y}\sum_{q_z}^{c_z+d_z}
E_{\bm{p}}^{\bm{a}\bm{b}} [\bm{p} \vert \bm{q}] E_{\bm{q}}^{\bm{c}\bm{d}},
\end{multline}
which we will express concisely as
\begin{equation}
\label{eq:MD def}
[\bm{a}\bm{b}\vert \bm{c}\bm{d}]=
\sum_{\bm{p}}^{\bm{a}+\bm{b}}\sum_{\bm{q}}^{\bm{c}+\bm{d}}
E_{\bm{p}}^{\bm{a}\bm{b}} [\bm{p} \vert \bm{q}] E_{\bm{q}}^{\bm{c}\bm{d}}.
\end{equation}
The intermediate tensors are defined recursively as
\begin{align}
E_{\bm{p}}^{\bm{a}^+ \bm{b}}&=\bm{p}^+ E_{\bm{p}^+}^{\bm{a}\bm{b}}+\bm{PA}\, E_{\bm{p}}^{\bm{a}\bm{b}}+\frac{1}{2\zeta} E_{\bm{p}^-}^{\bm{ab}}  \label{eq:EabP def1}\\
E_{\bm{p}}^{\bm{a}\bm{b}^+}&=\bm{p}^+ E_{\bm{p}^+}^{\bm{a}\bm{b}}+\bm{PB}\, E_{\bm{p}}^{\bm{a}\bm{b}}+\frac{1}{2\zeta} E_{\bm{p}^-}^{\bm{a}\bm{b}}  \label{eq:EabP def2}\\
E_{\bm{p}}^{\bm{0}\bm{0}}&=\begin{cases}
1, \text{ if } p=0\\
0, \text{ otherwise},
\end{cases} \label{eq:EabP def3}
\end{align}
and
\begin{equation}
    [\bm{p}\vert\bm{q}] = (-1)^q R_{\bm{p}+\bm{q}, 0}
\end{equation}
where $\bm{p} + \bm{q}$ is an component-wise vector addition and 
\begin{align}
    R_{\bm{p}^+,m}&=\bm{PQ}\, R_{\bm{p},m+1}+\bm{p} R_{\bm{p}^-,m+1}\\
R_{\bm{0},m}&=U_P U_Q (-2\theta)^m F_m (|\theta^2\bm{PQ}|^2) 
\end{align}
using the definitions from Eq.~\eqref{eq:boys-fn-def}.
Here we have used the compact notation discussed in Section \ref{sec:notation}. 

\begin{figure*}
    \centering
    \includegraphics[width=\linewidth]{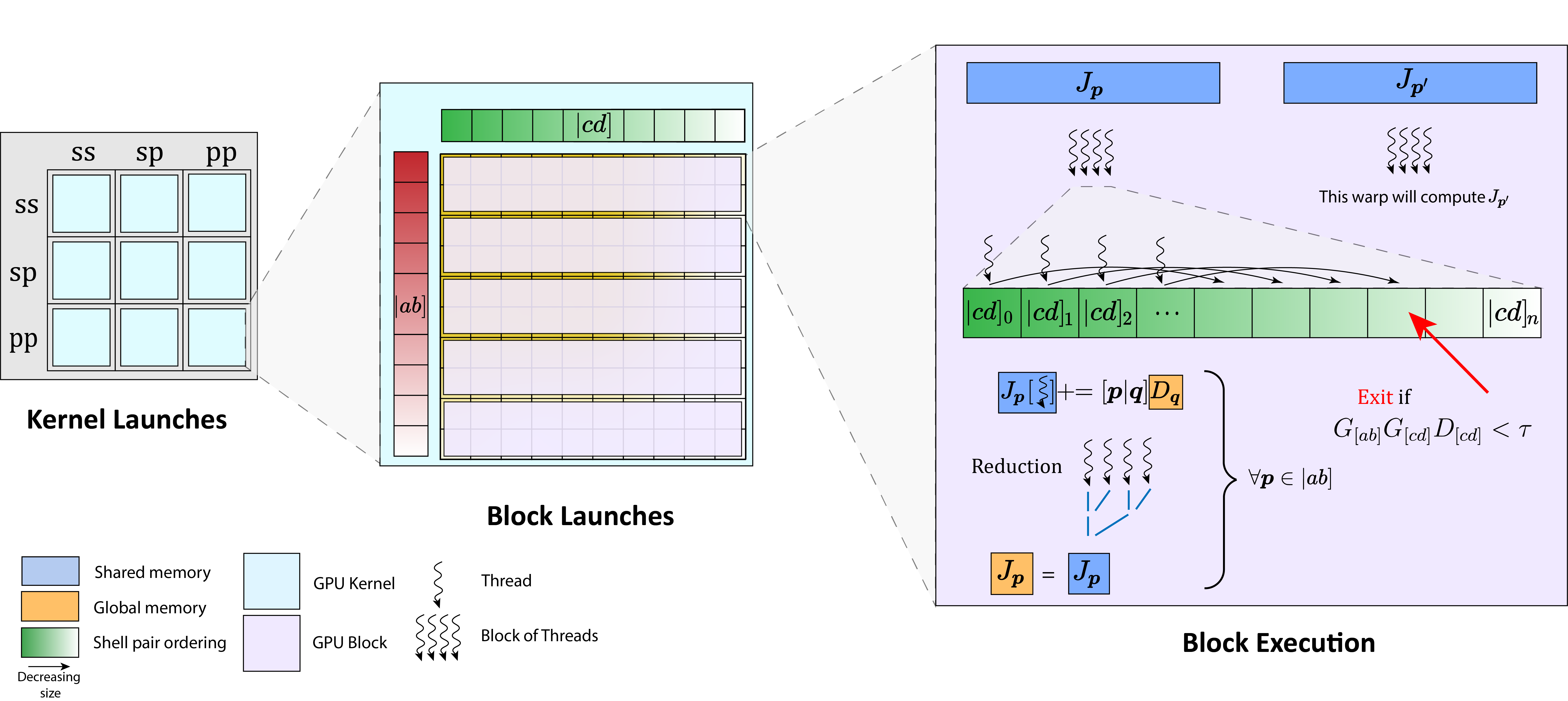}
    \caption{Algorithmic overview of the $\bm{J}$ matrix computation in our optimized UM09 scheme.}
    \label{fig:J}
\end{figure*}
As well as being a late contraction scheme, advantages of the MMD recurrence relations include precomputation of the $E_{\bm{p}}^{\bm{a} \bm{b}}$ which can occur once at the shell pair level prior to the SCF, and the very simple $[\bm p|\bm q]$ recurrence formulation.

A naive implementation of the MMD integral scheme is presented in Algorithm \ref{alg:naive MMD}.
There are significant performance concerns with this naive approach which will be detailed and rectified in Section \ref{sec:K_kernel_level}.

\begin{algorithm}[!htb]
    \caption{Naive MMD scheme implementation for the evaluating all integrals in $(ab\vert cd)$. Note that we are using the bounds of the for loops to indicate complexity, the exact angular momentum vectors are arduous to specify and their ordering is not important for this illustration.} \label{alg:naive MMD}
    $p\gets a+b$ \\
    $q\gets c+d$ \\
    \For{$\bm{p} \gets 1 \ \text{to}\ \frac{(p+1)(p+2)(p+3)}{6}$}{
    \For{$\bm{q} \gets 1 \ \text{to}\ \frac{(q+1)(q+2)(q+3)}{6}$}{
    Compute $[\bm{p}\vert \bm{q}]$ \\
    \For{$\bm{a} \gets 1 \ \text{to}\ \frac{(a+1)(a+2)}{2}$}{
    \For{$\bm{b} \gets 1 \ \text{to}\ \frac{(b+1)(b+2)}{2}$}{
    \For{$\bm{c} \gets 1 \ \text{to}\ \frac{(c+1)(c+2)}{2}$}{
    \For{$\bm{d} \gets 1 \ \text{to}\ \frac{(d+1)(d+2)}{2}$}{
        $[\bm{ab}\vert \bm{cd}] = E^{\bm{ab}}_{\bm{p}} [\bm{p}\vert \bm{q}] E^{\bm{cd}}_{\bm{q}}$ \\
        $F_{\bm{ab}}+=[\bm{ab}\vert \bm{cd}] D_{\bm{cd}}$ \\
        }
        }
        }
        }
        }
    }
\end{algorithm}
\subsection{\textbf{J} matrix implementation}

Our optimised GPU implementation of the construction of the $\bm{J}$ matrix is detailed in Fig.~\ref{fig:J}. 
In addition to the screening advantages of splitting $\bm{J}$ and $\bm{K}$, the MMD scheme also allows for pre- and post-contraction at the shell pair level for the computation of $\bm{J}$
\begin{align}\label{eq:pre and post contraction}
J_{[\bm{a}\bm{b}]} &=
\sum_{\bm{c}\bm{d}}\sum_{\bm{p}}^{\bm{a}+\bm{b}}\sum_{\bm{q}}^{\bm{c}+\bm{d}}
E_{\bm{p}}^{\bm{a}\bm{b}} [\bm{p} \vert \bm{q}] E_{\bm{q}}^{\bm{c}\bm{d}} D_{\bm{c}\bm{d}}\\
&=\sum_{\bm{p}}^{\bm{a}+\bm{b}}\sum_{\bm{q}}^{\bm{c}+\bm{d}}
E_{\bm{p}}^{\bm{a}\bm{b}} [\bm{p} \vert \bm{q}] D_{\bm{q}}\\
&=\sum_{\bm{p}}^{\bm{a}+\bm{b}}
E_{\bm{p}}^{\bm{a}\bm{b}} J_{\bm{p}}
\end{align}

Here we use the $J_{[\bm{a}\bm{b}]}$ notation to denote a primitive element of $\bm{J}$. The only contraction computed at the shell quartet level is $J_{\bm{p}} = \sum_{\bm{q}}[\bm{p} \vert \bm{q}] D_{\bm{q}}$. This reduces register usage, decreasing local memory reads and writes or increasing kernel occupancy. Custom kernels are auto-generated for each angular momentum class for efficient evaluation.

The formation of $D_{\bm{q}}$ is performed on GPU using a strided batched BLAS call on the primitive density matrix $D_{[\bm{c}\bm{d}]}$ and the $E_{\bm{q}}^{\bm{c}\bm{d}}$ tensor, with a separate BLAS call required for each angular momentum class. $D_{\bm{q}}$ is transposed before kernel launches to ensure contiguous reads.

The $J_{[\bm{a}\bm{b}]} =\sum_{\bm{p}} E_{\bm{p}}^{\bm{a}\bm{b}} J_{\bm{p}}$ contraction is similarly performed on GPU using a strided batched BLAS call per angular momentum class. Primitive $J_{[\bm{a}\bm{b}]}$ elements are then contracted into their final version $J_{\bm{a}\bm{b}}$ on the GPU using a customized kernel.

For computing $J_{\bm{p}}$, we adopt the method developed by \citeauthor{ufimtsev_quantum_2009} to remove atomics and for large system screening.\cite{ufimtsev_quantum_2009} Permutational symmetries $[\bm{ab}| \leftrightarrow [\bm{ba}|$ and $|\bm{cd}] \leftrightarrow |\bm{dc}]$ are exploited by pruning the shell pairs accordingly. The $|\bm{ab}]$ shell pairs are sorted by their Cauchy-Schwartz factor $G_{[ab]}$, while the $|\bm{cd}]$ shell pairs are sorted by the product of $G_{[cd]}$ and the maximum absolute element of the corresponding density matrix block. This sorting enables the algorithm to halt once the product of the $|\bm{ab}]$ and $|\bm{cd}]$ bounds falls below the screening threshold.

Figure \ref{fig:J} further illustrates our implementation. Each warp of threads is assigned an $[ab|$ shell pair and loops through the $|cd]$ shell pairs. Each thread computes all primitive integrals $[\bm{p}|\bm{q}]$ arising from the shell quartet, integrating them with the pre-contracted density block $D_{\bm{q}}$. Threads are assigned to shell quartets rather than basis function quartets for efficient evaluation. 
\begin{figure*}[h!]
    \centering
    \includegraphics[width=\linewidth]{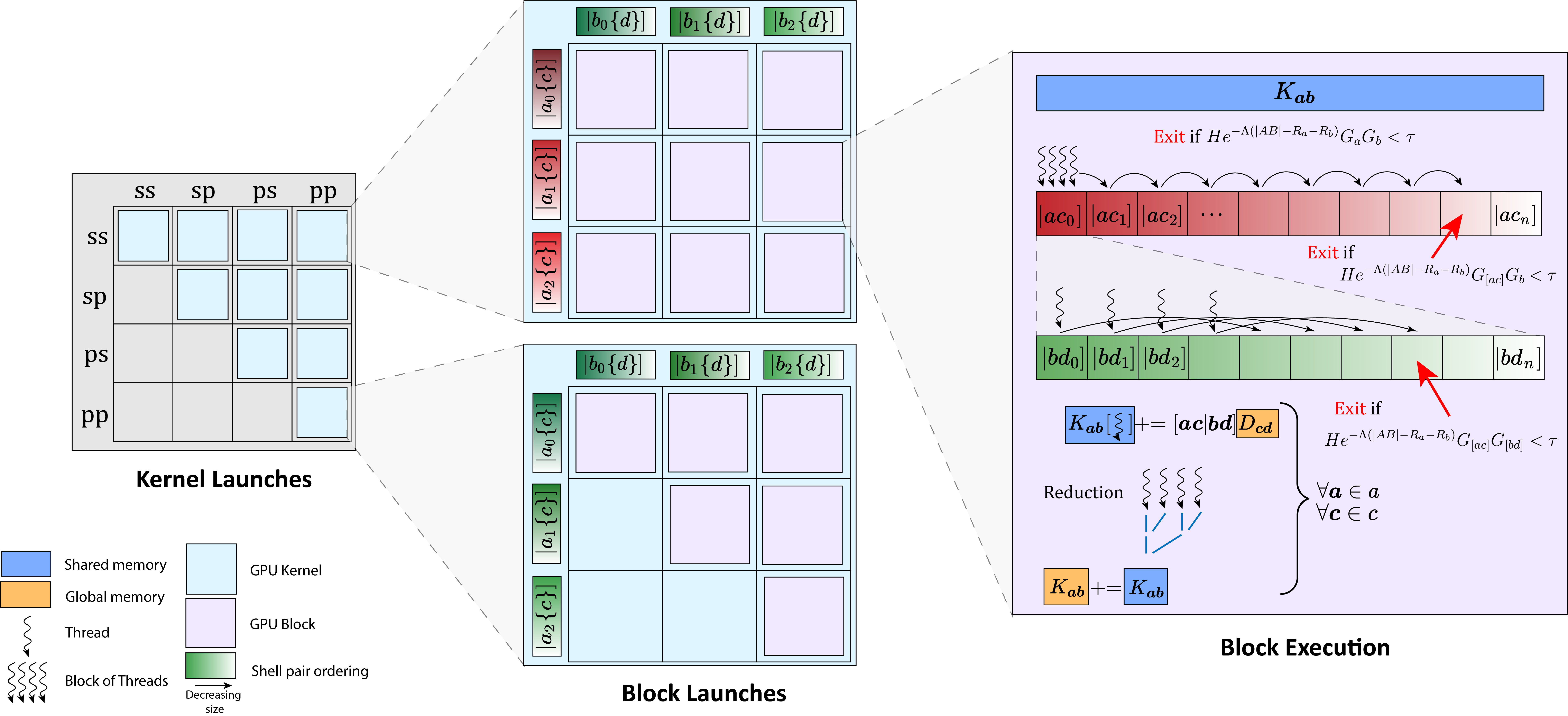}
    \caption{Algorithmic overview of the $\bm{K}$ matrix computation in our optimized UM09 scheme.}
    \label{fig:K}
\end{figure*}
\begin{figure}[h!]
    \centering
    \begin{subfigure}[b]{\linewidth}
        \centering
        \includegraphics[width=1\linewidth]{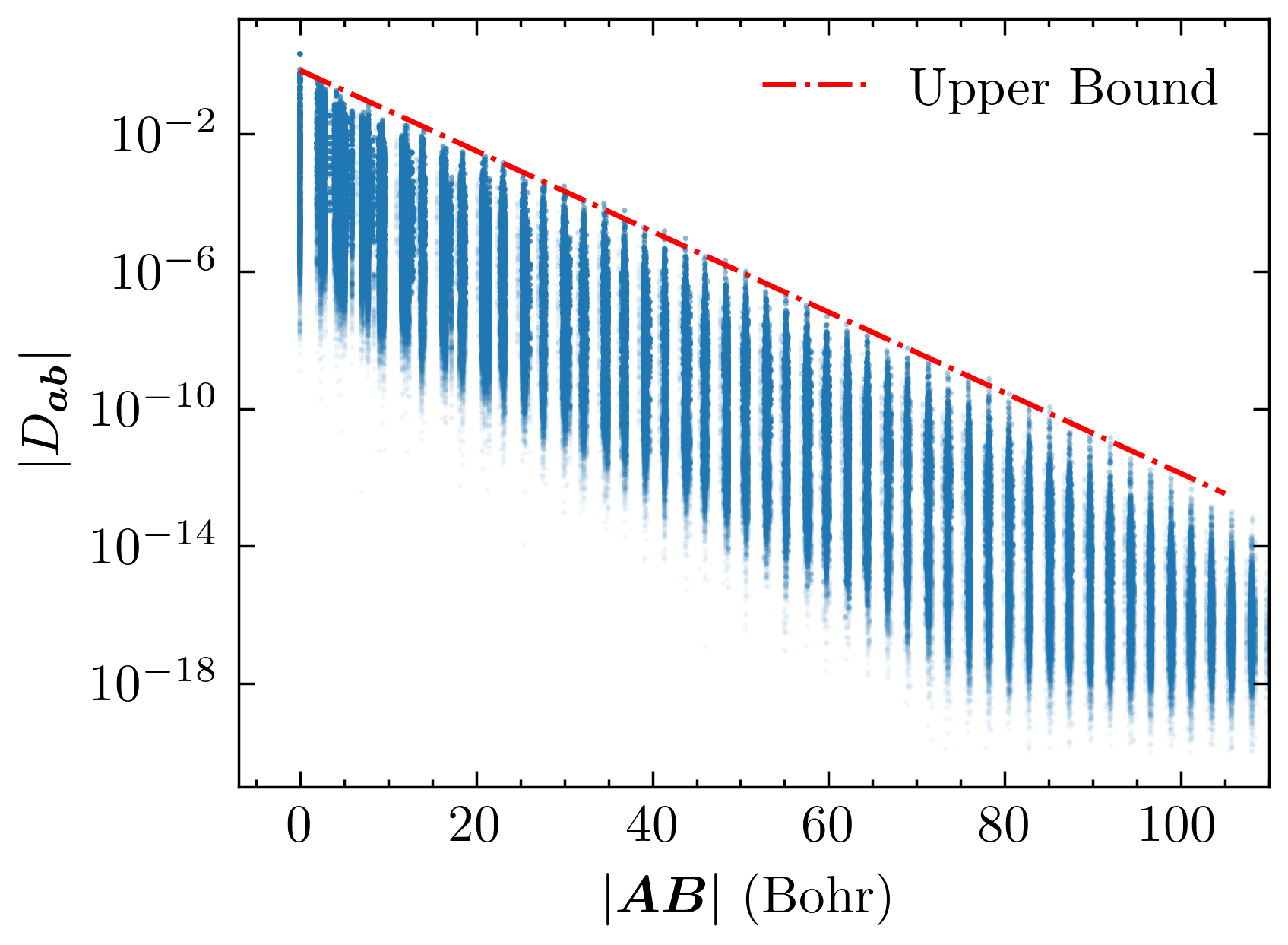}
        \caption{Bound for gly\textsubscript{20}}
        \label{fig:gly20_upper_bound}
    \end{subfigure}    
    \begin{subfigure}[b]{\linewidth}
        \centering
        \includegraphics[width=1\linewidth]{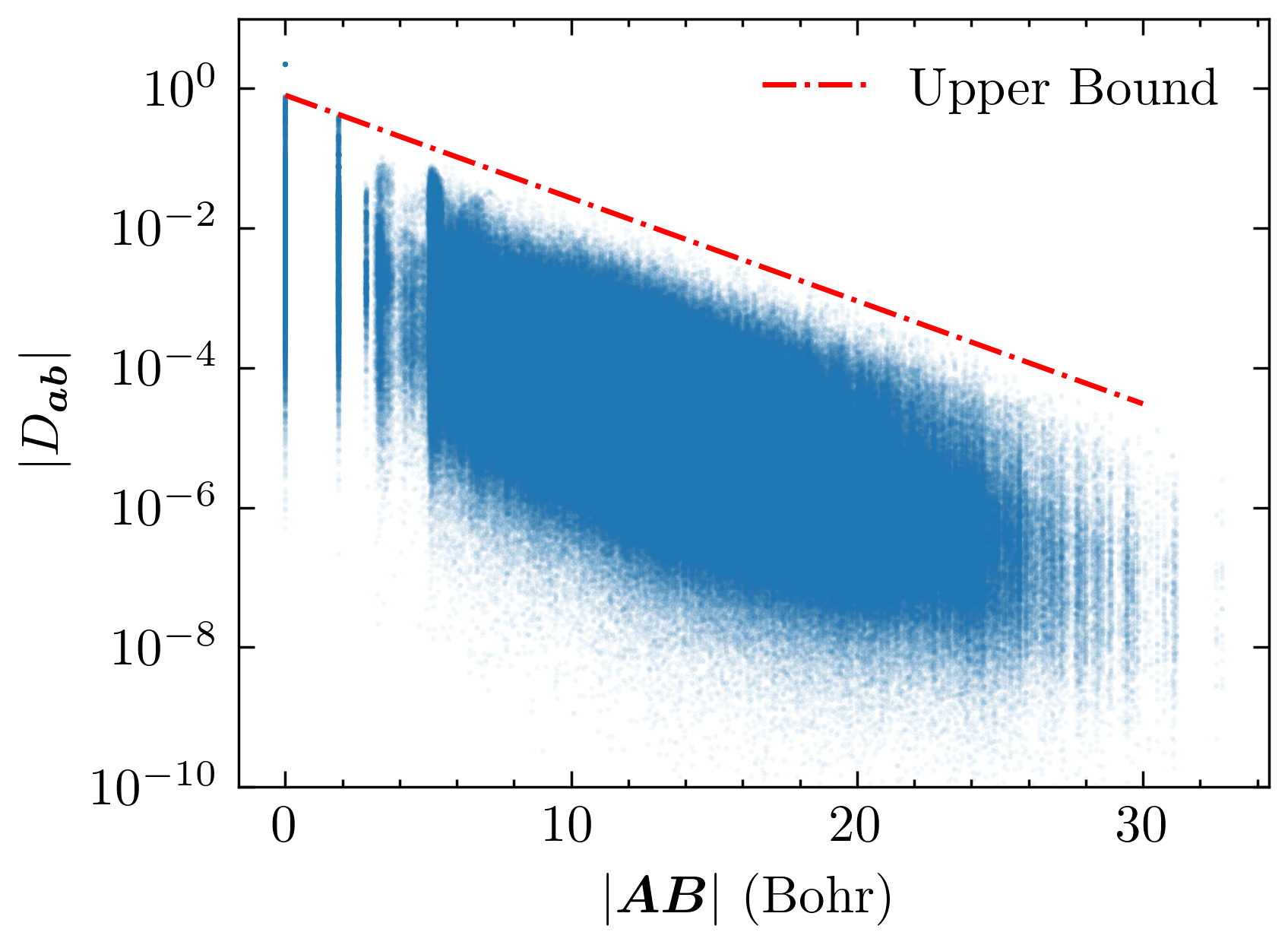}
        \caption{Bound for $(H_2O)$\textsubscript{60}}
        \label{fig:w60_upper_bound}
    \end{subfigure}
    
    \caption{Exponential decay with respect to distance between basis functions of the converged density matrix in gly\textsubscript{20} and $(H_2O)$\textsubscript{60} using the cc-pVDZ basis set.}
    \label{fig:upper-bounds}
\end{figure}
We monitor the Schwartz bound product and the maximum of the density block, halting the loop through $|cd]$ shell pairs when the product falls below the screening threshold. This ordering prevents warp divergence. A reduction over all threads in a warp is performed in shared memory to obtain the primitive $J_{\bm{p}}$ elements. A primitive $\bm{J}$ matrix is used to avoid atomic operations and facilitate post-contraction.

A warp per $[ab|$ ensures sufficient parallelism without bottlenecking reduction, minimizing warp divergence. Performance was maximized with a block dimension of 64, computing two $[ab|$ per block.

\subsubsection{Multi-GPU}
To distribute $\bm{J}$ computation across multiple GPUs, we dynamically distribute integral classes. Since a few integral classes dominate computation time, we split these classes into sub-batches using the strategy discussed in Section \ref{sec:HGP screening}. This detail is not depicted in Fig. \ref{fig:J} for simplicity, but it involves splitting the $ss$, $sp$, and $pp$ batches.

\subsection{\textbf{K} matrix implementation}
The GPU accelerated implementation of the $\bm{K}$ matrix formation has been designed to be performed in linear time complexity. We will first discuss how to achieve this while removing warp divergence and then discuss efficient integral computation using the MMD scheme.

\subsubsection{Integral class optimisation}
The algorithmic overview of our optimized approach for constructing $\bm{K}$ is shown in Fig. \ref{fig:K}.

Similar to the original UM06 algorithm, to facilitate fine grained density screening and to reduce atomic operations, the $\bm{K}$ implementation only exploits the $[ab|cd] \leftrightarrow [cd|ab]$ symmetry.

We construct batches of primitive shell pairs based on their angular momentum class, ordering them first by the initial shell index and then by their primitive shell pair Cauchy-Schwarz factor. In Fig. \ref{fig:K}, this sorting is represented using the notation $|a_0 \{c\}]$, which indicates a list of significant shell pairs $|a_0c_i]$ for a fixed shell $|a_0]$ and all $|c_i]$ in batch of shells $\{c\}$.

As depicted in Fig. \ref{fig:K}, each block of threads is assigned to the calculation of a single contracted shell pair block $K_{ab}$ of the exchange matrix. All threads within a block start with the same $[ac|$ shell pair, with each thread assigned a unique $|bd]$. Each thread evaluates the contribution to the exchange matrix from all basis functions in the shell quartet, incrementing a local, thread-private $K_{ab}$ buffer. The threads then proceed through the list of $|bd]$ shell pairs. Once all $|bd]$ pairs are processed, all threads move to the next $[ac|$ shell pair and continue to update their $K_{ab}$ buffer.

After computing all required integrals, the local $K_{ab}$ values are first reduced in shared memory, and then the resulting block is used to update the exchange matrix in global memory. We found that performance was maximized with a block dimension of 128. However, this block dimension must be reduced for high angular momentum kernels, as detailed in Section \ref{sec:f functions}.

Unlike the $\bm{J}$ scheme, we cannot order our shell pairs based on the magnitude of the density matrix, preventing us from determining when to terminate the scan through shell pairs. To address this, we utilize the exponential decay property of the density matrix, as established in previous studies\cite{kohn_analytic_1959, he_exponential_2001, maslen_locality_1998, ismail-beigi_locality_1999, barca_q-mp2-os_2020, goedecker_decay_1998}, to construct the following upper bound:
\begin{equation}
\label{eq:density_bound}
   |D_{cd}|\le H\, e^{-\Lambda|\bm{CD}|}.
\end{equation}
Here, $H$ and $\Lambda$  are system-dependent parameters, which are determined at each SCF iteration, as discussed later in this Section. The non-bold notation $D_{cd}$ is used to indicate that all density elements  $D_{\bm{cd}}$ within the shell pair block $|cd]$ are bounded by this expression. We use similar notation to bound an entire shell pair block of $\bm{K}$ below.

Figures \ref{fig:gly20_upper_bound} and \ref{fig:w60_upper_bound} show the upper bound and the magnitude of the density matrix elements plotted against the distance between basis function centers for a polyglycine system with 20 monomer units and a 60-water cluster, respectively, both using the cc-pVDZ basis set. The exponential decay is evident in both cases.

To bound a primitive block of $K_{[ab]}$ by utilizing Eq.~\eqref{eq:density_bound}, we proceed as follows. We define:
\begin{align}
    R_{a} &= \max_{c} \left( |\bm{AC}|: G_{[ac]} < \tau \right) \\
    G_{a} &= \max_{c} \left( G_{[ac]} \right)
\end{align}
It can be easily shown that:
\begin{align}
    |\bm{CD}| &\ge |\bm{AB}| - R_{a} - R_{b}.
\end{align}
Thus, each contribution $D_{cd}[ac|bd]$ to $K_{[\bm{ab}]}$ can be bounded as follows:
\begin{align}
    |D_{cd} [ac|bd]| &\le H\, e^{-\Lambda|\bm{CD}|} G_{[ac]} G_{[bd]} \\
    &\le H\, e^{-\Lambda(|\bm{AB}| - R_{a} - R_{b}) }G_{a} G_{b}
\end{align}
Therefore, we have:
\begin{align}
    |K_{[ab]}| &\le \sum_{cd} |D_{cd} [ac|bd]| \\
    &\le N_{cd}^{ab}\, H\, e^{-\Lambda(|\bm{AB}| - R_{a} - R_{b})} G_{a} G_{b}.
\end{align}
Here, $N_{cd}^{ab}$ represents the number of shell pairs $|cd]$ that yield numerically significant contributions 
$|D_{cd} [ac|bd]|$ when coupled with $|ab]$. Due to the combined exponential decay of the density matrix and the Gaussian nature of the PGFs, there are only $\mathcal{O}(1)$ shell pairs $|cd]$ that result in a significant product $|D_{cd} [ac|bd]|$ for each $|ab]$. This implies that $N_{cd}^{ab}$ is bounded by a system-dependent constant $N^{*}$, 
\emph{i.e.}, $N_{cd}^{ab} \le N^{*}$ for all $|cd]$ and $|ab]$.

This method allows us to rigorously determine whether an element of $\bm{K}$ is insignificant using only the data associated with $a$ and $b$. Each block shares $a$ and $b$, enabling us to assess the insignificance of $K_{[ab]}$
before evaluating a single integral, by checking whether 
\begin{equation}
    H\, e^{-\Lambda(|\bm{AB}| - R_{a} - R_{b})} G_{a} G_{b} \le \tau^{*}
\end{equation}
where $\tau^{*} = \tau/N^{*}$, and $\tau$ is a user-define threshold. Consequently, if insignificance is established, the $K_{[ab]}$ block can be skipped.

To compute $H$ and $\Lambda$, we bin the elements of $D_{\bm{cd}}$ by $|\bm{CD}|$, take the maximum absolute value in each bin, and perform a least squares linear regression on the logarithm of these values. These parameters must be recomputed at each SCF iteration when the density matrix $\bm{D}$ is updated. Note that we exclude density elements from basis functions centered on the same atom, as this skews the exponential decay. This exclusion is valid since these density elements will not be screened regardless.

We can also use tighter variations of this bound within the $[ac|$ and $|bd]$ loops of the kernel. Once inside the $[ac|$ loop we can replace $G_{a}$ with the much tighter upper bound $G_{[ac]}$. Similarly we can tighten $G_{b}$ to $G_{[bd]}$ once inside the $|bd]$ loop. As shell pairs are sorted by their Cauchy Schwartz bound all threads will exit the loop at the same time preventing warp divergence. 

It is worth comparing this approach to that discussed by \citeauthor{luehr_gaussian_2016}.\cite{luehr_gaussian_2016}
They suggest rigorously bounding $\bm{K}$ elements as follows
\begin{equation}
    K_{[\bm{ab}]} \le (\bm{GDG})_{[\bm{ab}]}
\end{equation}
where $\bm{G}$ is a matrix of primitive Cauchy Schwarz bounds. While this is a tighter bound for $K_{[\bm{ab}]}$ as it uses the exact values of the density matrix, it is not possible to tighten this bound once inside the $[ac|$ or $|bd]$ loops. We observed significant performance benefits from bound tightening within these loops particularly on globular systems. Additionally, the computation of $\bm{GDG}$ is significantly more expensive than finding an upper bound for elements of $\bm{D}$. Hence, we did not attempt to implement this strategy, however using both bounds in conjunction could be worth considering in future work.

Experimentally we found that ignoring elements of $\bm{K}$ whose magnitude is less than $\tau^{*} =5* 10^{-6}$ does not change the final HF energy by more than $10^{-5}E_h$. This implies our $\bm{K}$ screening threshold can be significantly looser than the $\tau = 10^{-10}$ threshold required for the $\bm{J}$ and Brc21 schemes. 

\subsubsection{Per integral optimisation}\label{sec:K_kernel_level}

Let us now discuss how to efficiently compute 
\begin{equation}
 [\bm{a}\bm{c}\vert \bm{b}\bm{d}]\,D_{\bm{c}\bm{d}} =
\sum_{\bm{p}}^{\bm{a}+\bm{c}}\sum_{\bm{q}}^{\bm{b}+\bm{d}}
E_{\bm{p}}^{\bm{a}\bm{c}} [\bm{p} \vert \bm{q}] E_{\bm{q}}^{\bm{b}\bm{d}} D_{\bm{c}\bm{d}} 
\end{equation}
on GPU.

An important consequence of Eqs.~\eqref{eq:EabP def1}-\eqref{eq:EabP def3} is that
\begin{equation}\label{eq:MMD-observation}
    E^{\bm{ab}}_{\bm{a+b}} = \frac{1}{(2\zeta)^{a+b}}.
\end{equation}
Due to the simplicity of the computation of $1/(2\zeta)^{a+b}$ it is more efficient in terms of both register usage and reducing global reads to compute $E^{\bm{ab}}_{\bm{a+b}}$ in the GPU kernel rather than reading it in.

The remaining elements of $E^{\bm{ab}}_{\bm{p}}$ are computed once on CPU before the first SCF iteration and read from global memory. It is ensured that these or ordered such that reads are coalesced.  

The dimensions of an $E^{\bm{ab}}_{\bm{p}}$ tensor for a $\vert ab]$ shell pair are $\left(\frac{(a+1)(a+2)}{2} , \frac{(b+1)(b+2)}{2}, \frac{(a+b+1)(a+b+2)(a+b+3)}{6}\right)$. However, for a given PGF pair $\vert \bm{ab}]$ there are only $(\bm{a}_x + \bm{b}_x+1)(\bm{a}_y + \bm{b}_y+1)(\bm{a}_z + \bm{b}_z+1)$ non zero $E^{\bm{ab}}_{\bm{p}}$ elements, by the bounds of Eq.~\eqref{eq:MD def}. For example, for a $\vert dd]$ shell pair, there are 1{,}260 elements in the corresponding $E^{\bm{ab}}_{\bm{p}}$ tensor however only 336 of these are non zero.
% Check:
% Therefore, exploiting sparsity in a $(dd\vert dd)$ integral class may result in a 14 times FLOP reduction.

\begin{algorithm}[!htb]
    \caption{Example of array being forced into local GPU memory \label{alg:local mem}}
    array[10] \\
    $s\gets 0$ \\
    \#pragma unroll 1 \\
    \For{$i \gets 1$ to $10$}{
        s += array[i]
    }
\end{algorithm}

The final performance concern is more subtle. 
Consider Algorithm \ref{alg:local mem} in which each GPU thread computes a sum of a local array.
It is well known that reducing register usage per thread is required for high GPU occupancy due to limited GPU memory resources. It is thus intuitive to prevent the compiler manually unrolling the loop as this will increase register usage. However, the \texttt{\#prama unroll 1} command forces the compiler to write the array to local memory in order to index it.
The high latency associated with GPU local memory accesses makes this behaviour extremely undesirable.
It is only when this loop is fully unrolled that the values of the array do not need to be written to local memory.

Manually unrolling loops allows us to address each of these concerns.
We start by contracting $[\bm{p}\vert \bm{q}]$ with $E^{\bm{bd}}_{\bm{q}}$ for a given $\vert \bm{bd}]$ and all $\bm{p}$. We can then contract this unrolled tensor with $E^{\bm{ac}}_{\bm{q}}$ to construct ERIs for all $\vert \bm{ac}]$. As we compute these ERIs we immediately digest them with the density matrix into the private Fock buffers to reduce live registers. To reduce global reads from the density matrix we write each required density block into a private thread buffer.
Importantly, in each of these contractions we only contract with the non zero elements of $E^{\bm{ac}}_{\bm{p}}$ and $E^{\bm{bd}}_{\bm{q}}$.

The manual loop unrolling was written with code generators. We also note that auto generated manual loop unrolling is used to optimise the contraction within the $\bm{J}$ kernels as well.

\subsubsection{Multi-GPU}

To efficiently distribute the computation of $\bm{K}$ across multiple GPUs, it is essential to divide integral class batches to prevent any GPU from remaining idle. This process is algorithmically straightforward. Given that the batches are already sorted by shell index, they can be split at shell index boundaries. We employ similar heuristics to those discussed in Section \ref{sec:HGP screening} to determine the appropriate batch sizes.

\subsection{Extension to $f$ functions} \label{sec:f functions}
The challenges of extending our Fock build schemes to $f$-functions arise from the significantly increased complexity of the recurrence relations. While the MMD recurrence relations are considerably simpler than the HGP ones, the need for loop unrolling makes extending to high angular momentum equally challenging. This concern is specific to the computation in $\bm{K}$ kernels, as the 
$\bm{J}$ kernels are less complex due to pre- and post-contraction.

One potential strategy to reduce kernels' complexity is to assign threads to basis functions rather than shells. At high angular momentum, this approach can significantly distribute the workload (\emph{e.g.}, 10 ways for an $f$-shell), but it sacrifices the reuse of recursive intermediates. For evaluating $\bm{K}$, this can be managed by computing a single element of $K_{\bm{[ab]}}$ per block, rather than all elements in an $|ab]$ shell. This requires launching $\frac{(a+1)(a+2)}{2}\frac{(b+1)(b+2)}{2}$ kernels per integral class (\emph{e.g.}, 100 kernel launches for $(ff|ff)$) instead of just one. Implementing this scheme resulted in a minor performance decrease, so our $f$-function implementation does not involve splitting. However, this approach did reduce register spills, making it potentially beneficial for GPUs with limited memory.

If kernels are not split, the shared memory required to store the $K_{[ab]}$ and $J_{[ab]}$ blocks exceeds the limit of A100 GPUs. For instance, the number of doubles required to store $K_{[ab]}$ scales as $\frac{(a+1)(a+2)}{2} \frac{(b+1)(b+2)}{2} \verb|blockdim|$. Therefore, we simply reduce the block dimension \verb|blockdim| of these high angular momentum kernels, as the performance improvement from shared memory reduction significantly outweighs the reduction in block dimension.

%===============
\section{Code Generators}\label{sec:Code generators}
%===============
The nature of the recurrence relations required for efficient ERI evaluation results in a large number of shared recursive intermediates between integrals arising from a specific shell quartet. For example, in the HGP scheme all integrals in a given shell quartet have the same fundamental integrals $[\bm{00}|\bm{00}]^{m}$. 

Identifying an optimal evaluation strategy that minimizes the number of recursive intermediates presents a highly complex tree search problem. For the MMD method, exact solutions exist for total angular momentum $L = (a+b+c+d) \le 7$, while near-optimal heuristics are used for larger $L$ values~\cite{johnson_exact_1991}. Similarly, for the HGP method, some optimal solutions have been obtained through exhaustive search for VRR generated classes~\cite{liu_optimal_2016}. Near-optimal solutions can also be obtained using suitable heuristics for higher angular momentum integral classes~\cite{ryu_optimal_1993, johnson_efficient_1993}.

In our implementation, we employ heuristic approaches to achieve a near-optimal number of intermediates necessary for computing a given integral class. This approach addresses the tedious and error-prone task of manually generating recurrence relations, especially for high angular momentum integral classes, through the development of a dedicated code generator. The process begins with generating a Directed Acyclic Graph (DAG) of recursive intermediate dependencies specific to the adopted tree-search heuristics, followed by topologically sorting these intermediates. The topological sort is implemented using a priority queue, accommodating fully contracted, half-contracted, and uncontracted intermediates.

Additionally, the code generator produces a FLOP count for fully contracted, half-contracted, and uncontracted computations, facilitating the comparison between different evaluation schemes.

\begin{figure*}
    \centering
    \begin{subfigure}[b]{0.49\linewidth}
        \centering
        \includegraphics[width=\linewidth]{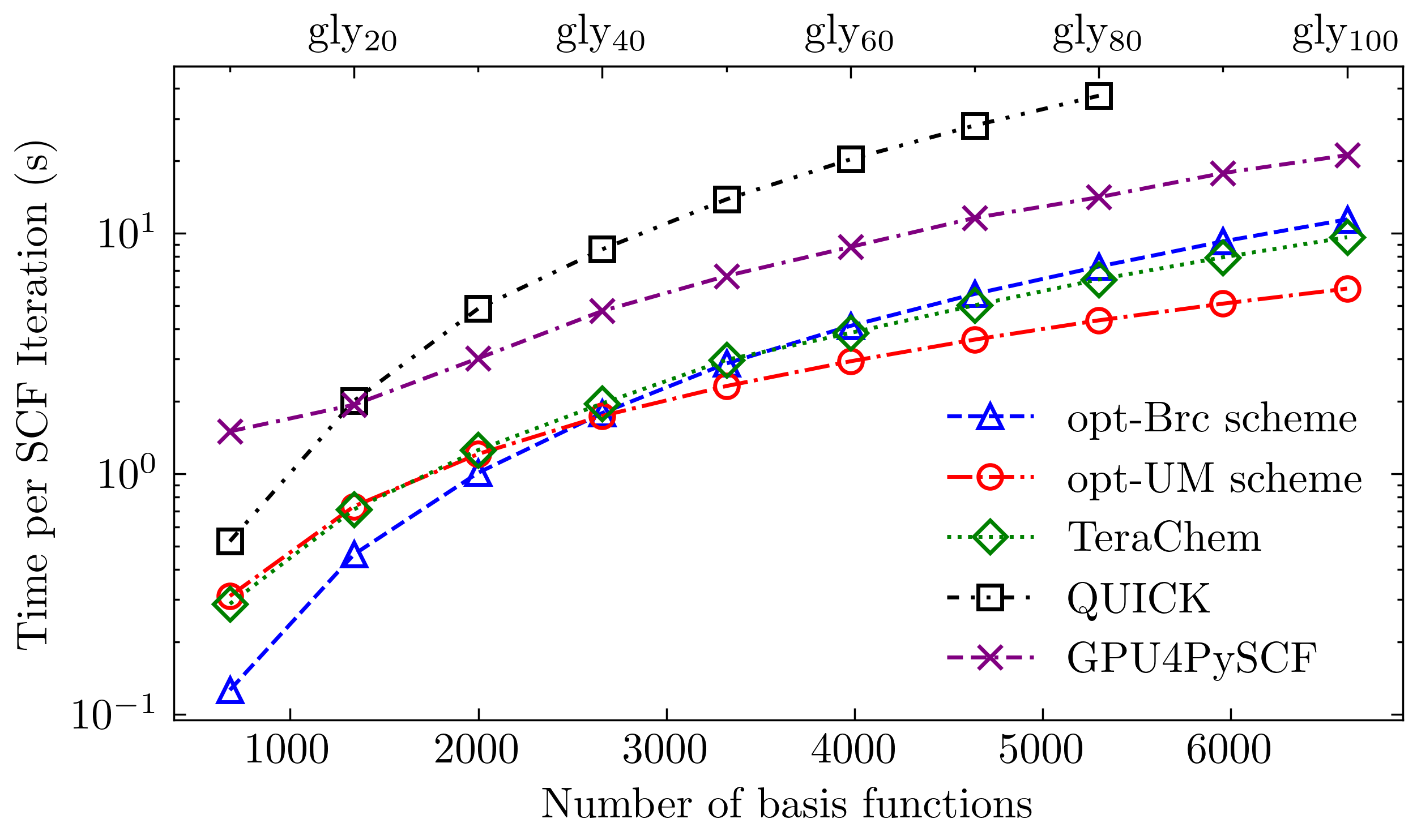}
        \caption{Varying length glycine chains with the 6-31G* basis}
        \label{fig:gly-6-31G*}
    \end{subfigure}    
    \hfill
    \begin{subfigure}[b]{0.49\linewidth}
        \centering
        \includegraphics[width=\linewidth]{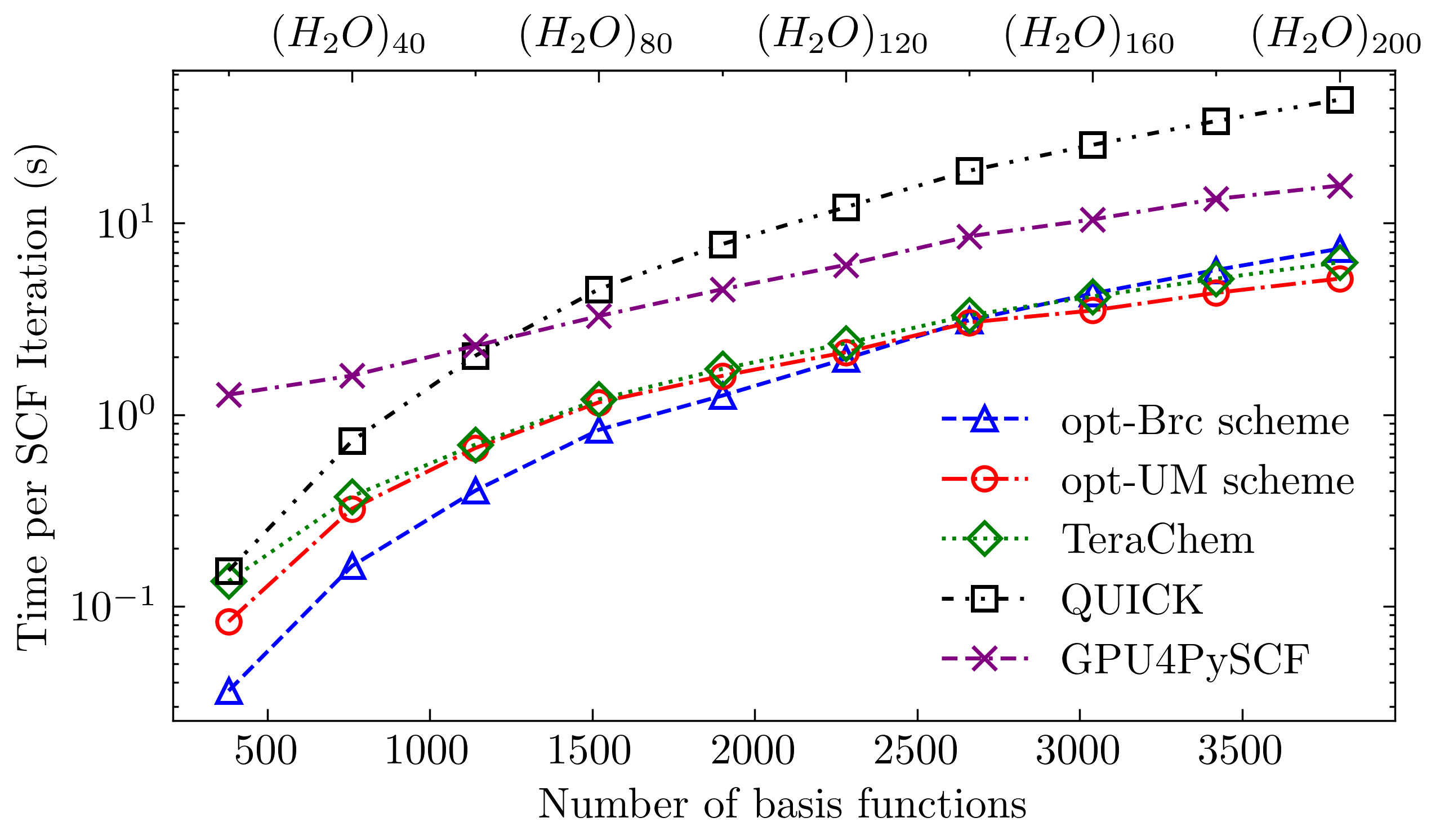}
        \caption{Varying size water clusters with the 6-31G* basis}
        \label{fig:water-6-31G*}
    \end{subfigure}
    \begin{subfigure}[b]{0.49\linewidth}
        \centering
        \includegraphics[width=\linewidth]{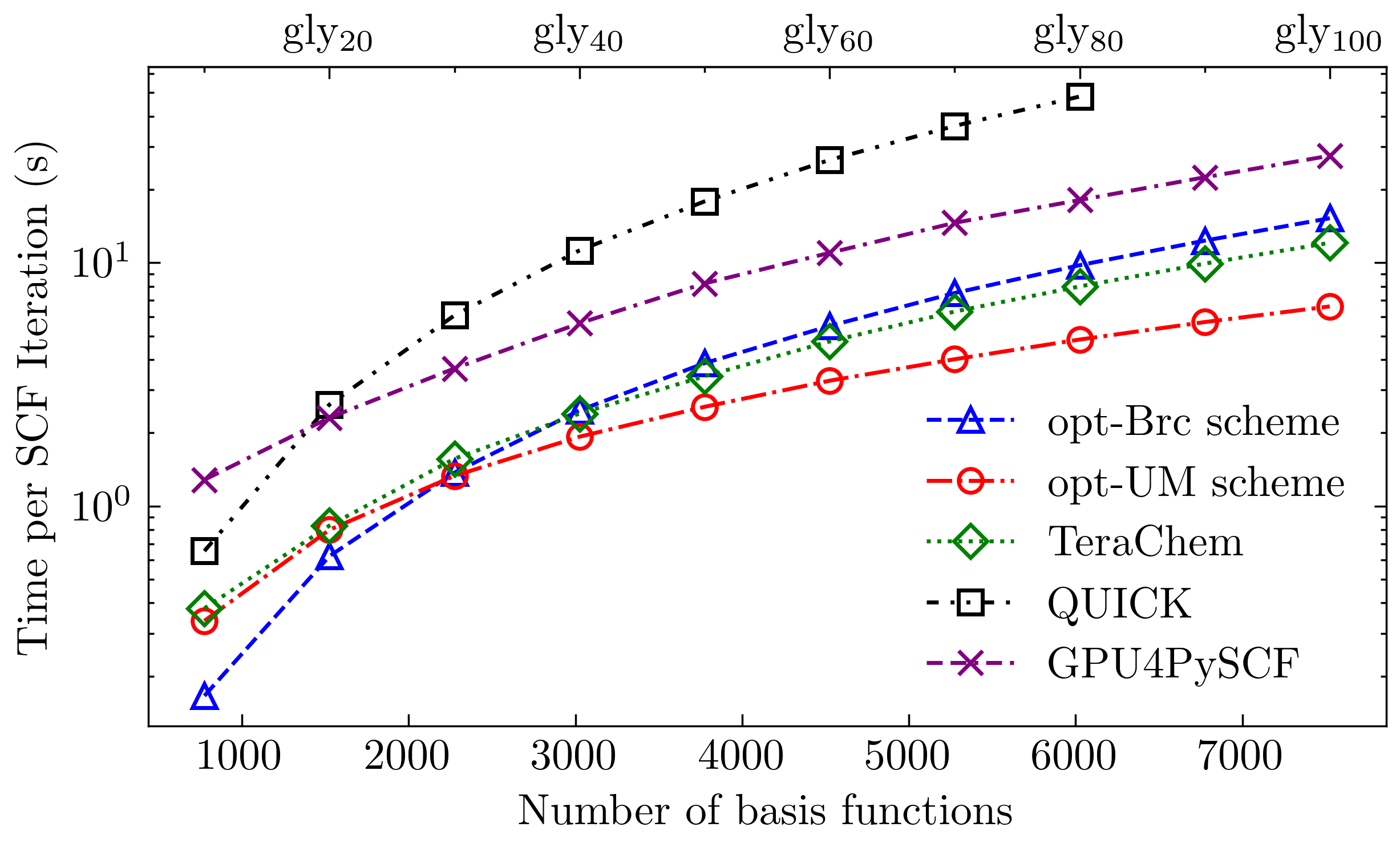}
        \caption{Varying length glycine chains with the 6-31G** basis}
        \label{fig:gly-6-31G**}
    \end{subfigure}    
    \hfill
    \begin{subfigure}[b]{0.49\linewidth}
        \centering
        \includegraphics[width=\linewidth]{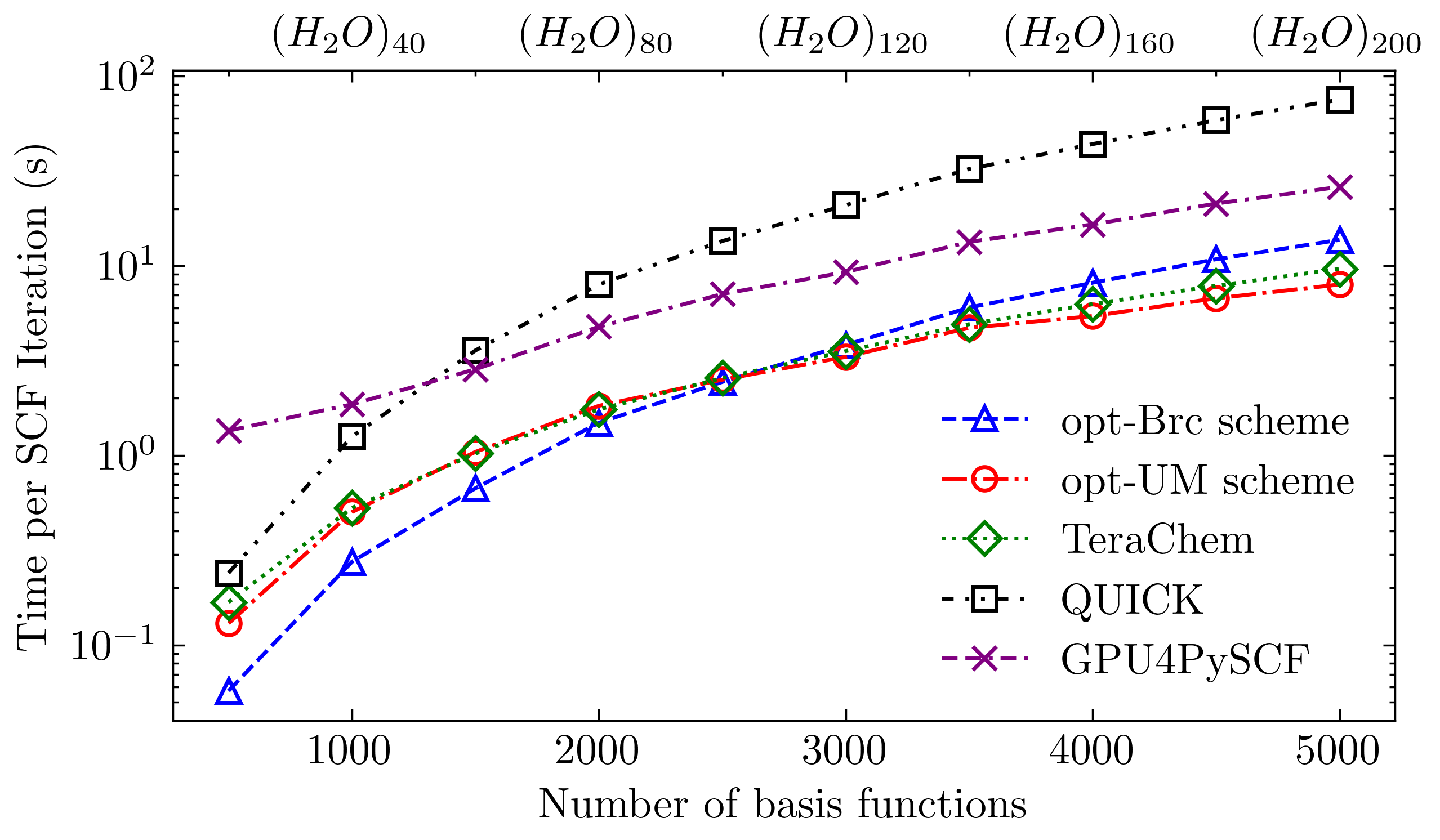}
        \caption{Varying size water clusters with the 6-31G** basis}
        \label{fig:water-6-31G**}
    \end{subfigure}
    \begin{subfigure}[b]{0.49\linewidth}
        \centering
        \includegraphics[width=\linewidth]{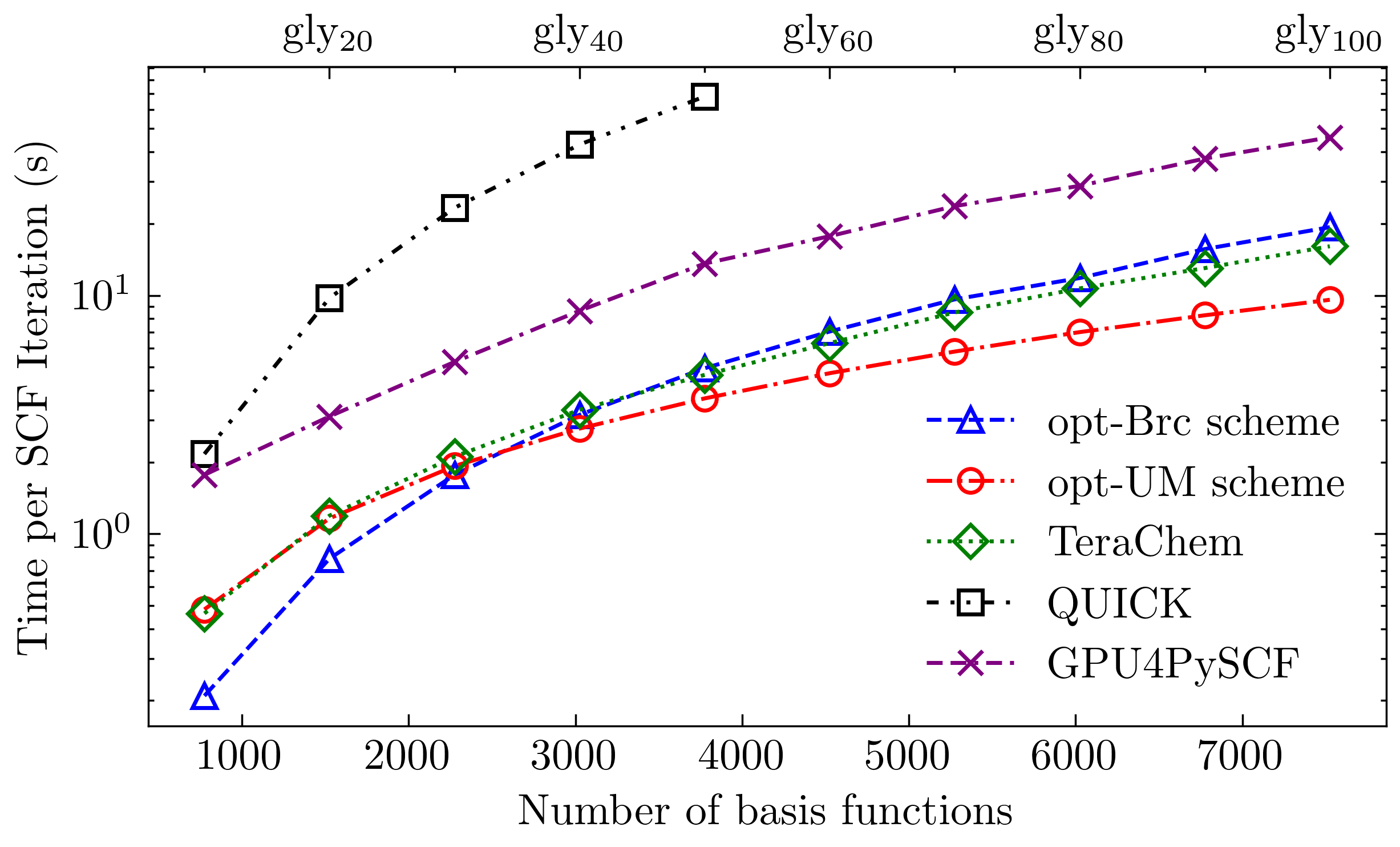}
        \caption{Varying length glycine chains with the cc-pVDZ basis}
        \label{fig:gly-cc-pVDZ}
    \end{subfigure}    
    \hfill
    \begin{subfigure}[b]{0.49\linewidth}
        \centering
        \includegraphics[width=\linewidth]{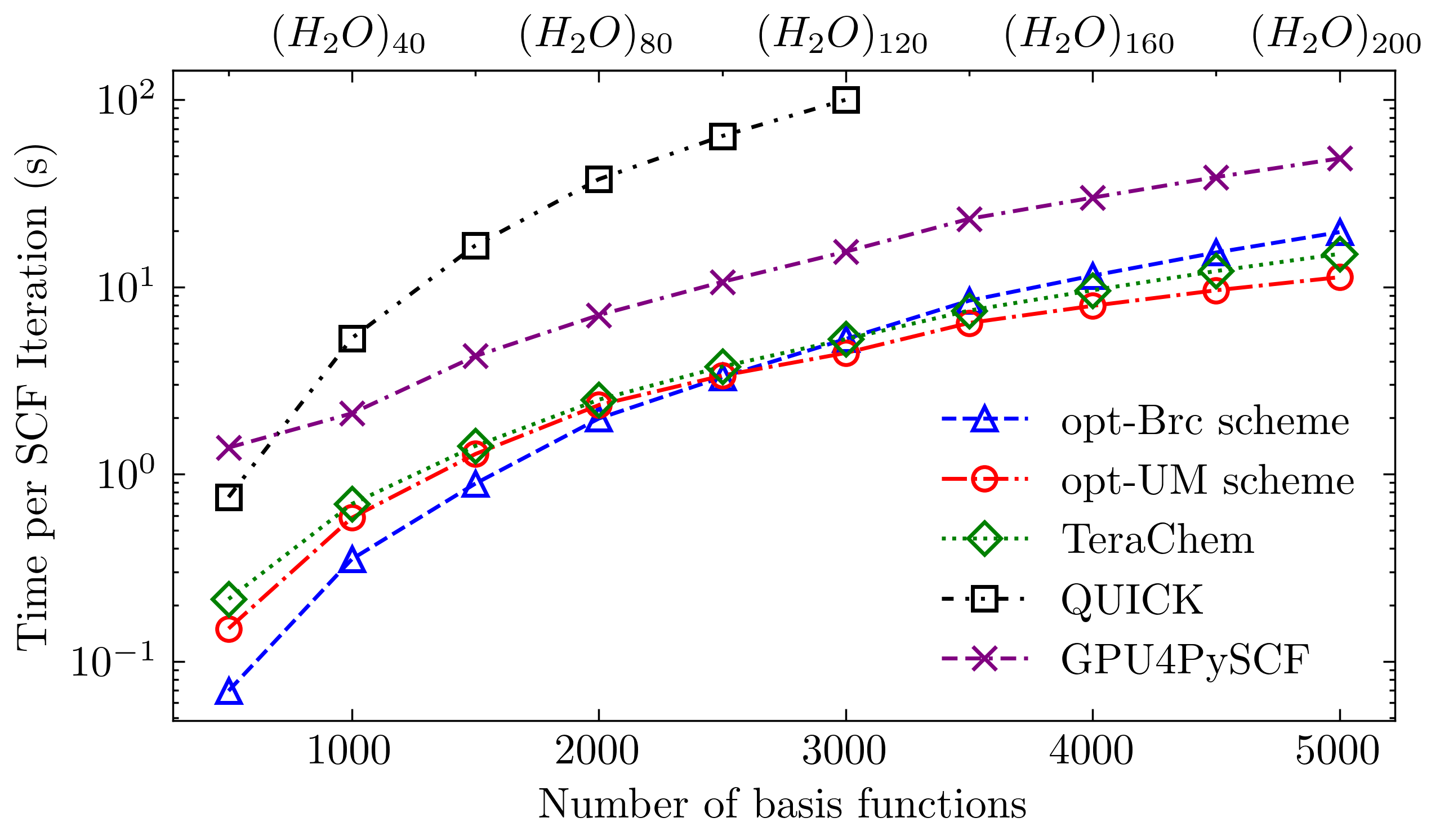}
        \caption{Varying size water clusters with the cc-pVDZ basis}
        \label{fig:water-cc-pVDZ}
    \end{subfigure}
    \begin{subfigure}[b]{0.49\linewidth}
        \centering
        \includegraphics[width=\linewidth]{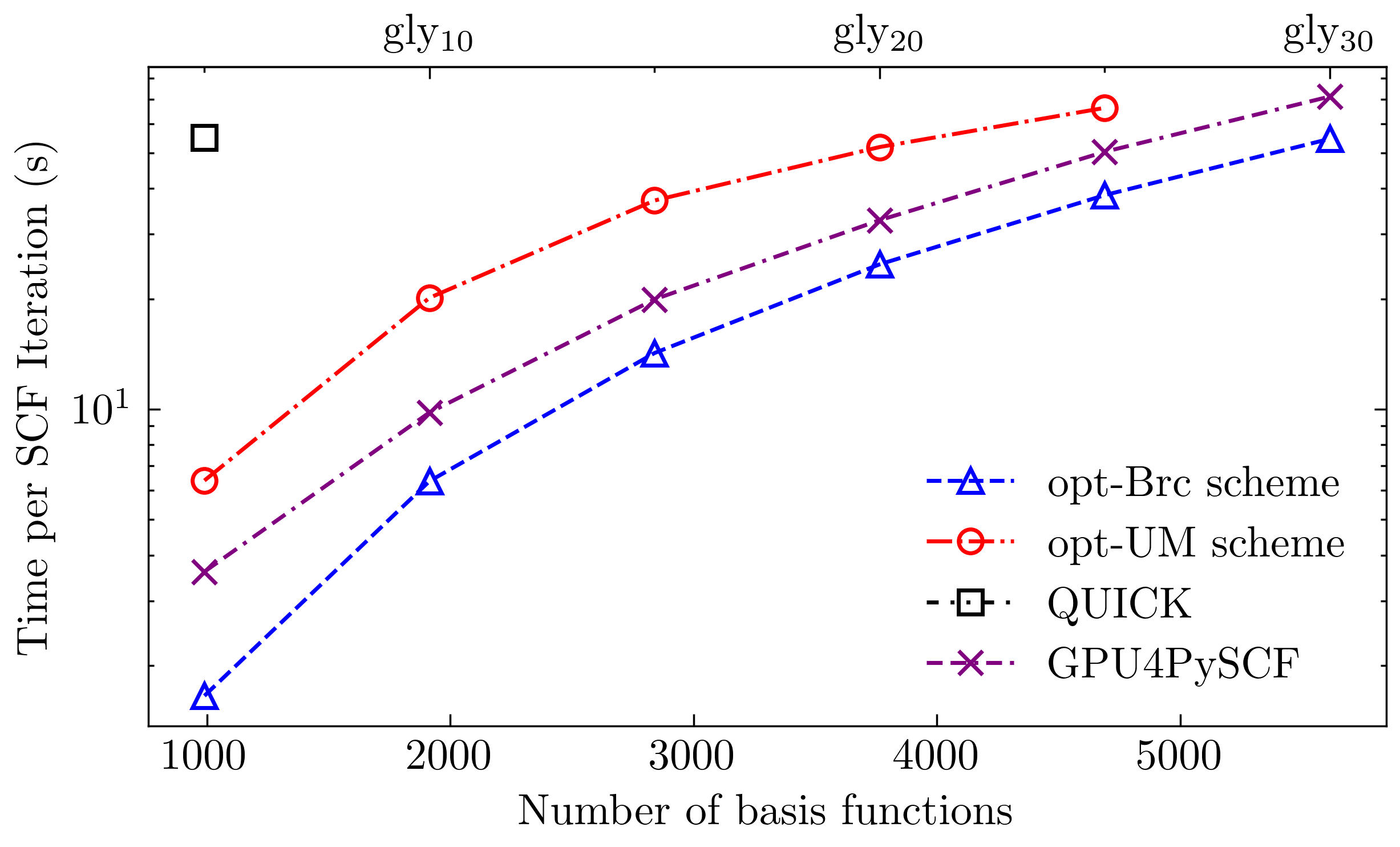}
        \caption{Varying length glycine chains with the cc-pVTZ basis}
        \label{fig:gly-cc-pVTZ}
    \end{subfigure}    
    \hfill
    \begin{subfigure}[b]{0.49\linewidth}
        \centering
        \includegraphics[width=\linewidth]{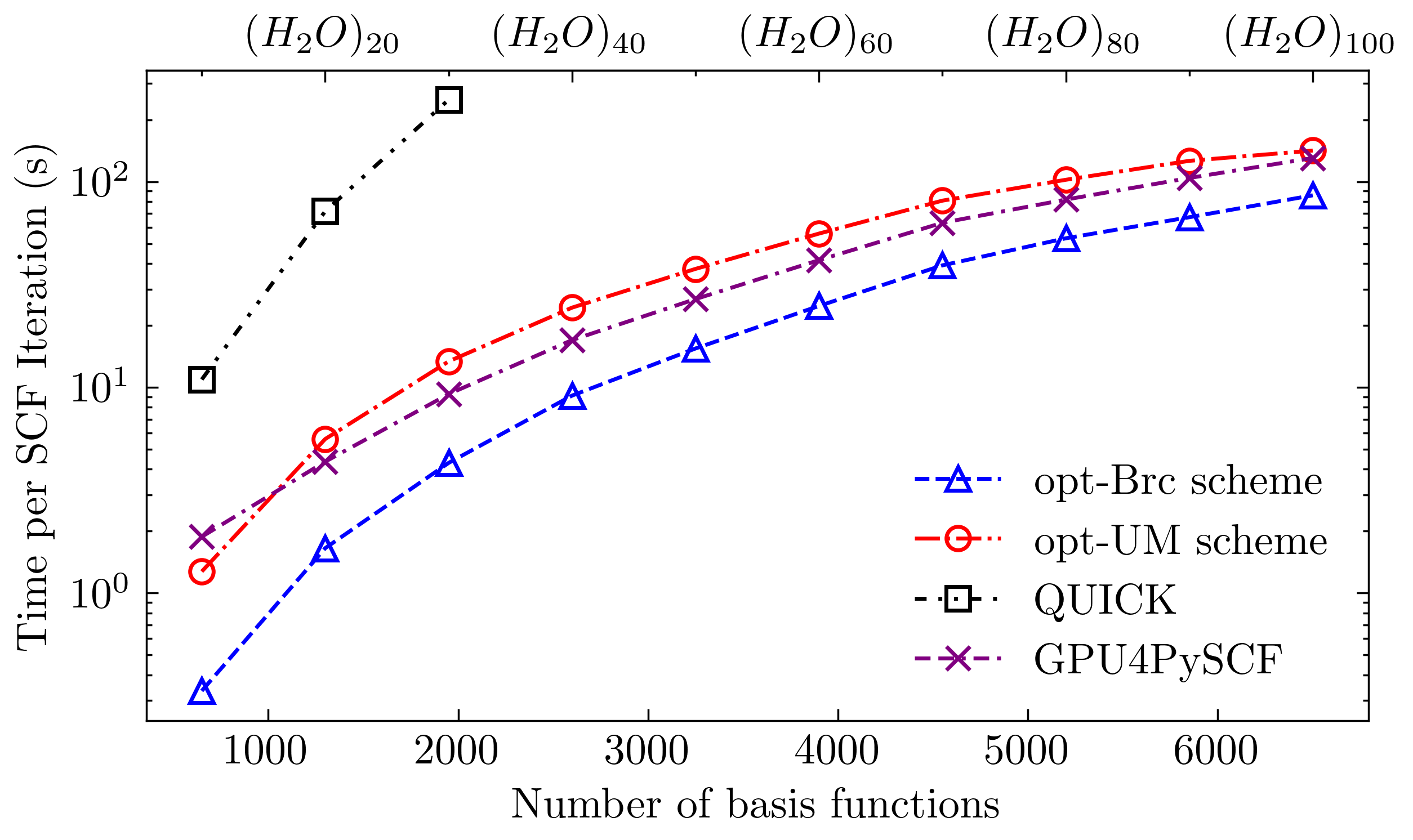}
        \caption{Varying size water clusters with the cc-pVTZ basis}
        \label{fig:water-cc-pVTZ}
    \end{subfigure}
    \caption{Single A100 GPU Fock build timings of linear and globular systems with varying basis sets of the opt-Brc and opt-UM schemes relative to implementations in \texttt{TeraChem}, \texttt{QUICK} and \texttt{GPU4PySCF}.}
    \label{fig:comparisons}
\end{figure*}
%===============
\section{\label{sec:results}Results}
%===============

In this Section, we present the performance results of our optimized GPU implementations. We refer to our optimized versions of the Brc21 and UM09 algorithms as opt-Brc and opt-UM, respectively. 

All results were obtained using a single node of the NERSC Perlmutter supercomputer, which is equipped with an AMD EPYC 7763 CPU and four NVIDIA A100 40GB GPUs. Tests using triple-$\zeta$ basis sets were conducted on Perlmutter nodes with 80GB GPUs to accommodate the substantial memory requirements of the CUDA runtime for handling high angular momentum kernel register spills.

We have elected to use a variety of common basis sets, in particular cc-pVDZ~\cite{dunning_gaussian_1989}, 6-31G*~\cite{hariharan_influence_1973} and 6-31G**~\cite{hariharan_influence_1973}, along with the triple-$\zeta$ basis set cc-pVTZ~\cite{dunning_gaussian_1989} to demonstrate the \textit{f} function capability. All basis sets were sourced from the Basis Set Exchange.~\cite{pritchard_new_2019}.

For performance benchmarking, we selected linear and globular systems of increasing sizes, focusing on polyglycine chains and water clusters consistent with our previous work~\cite{stocks_high-performance_2024}. These systems were chosen due to their scalability, allowing us to systematically observe scaling trends. Each glycine unit ($C_2 H_3 N O$) contributes 30 electrons across 7 atoms to a polyglycine chain of length $n$, which is denoted as $\text{gly}_n=H(C_2 H_3 N O)_n OH$. The water clusters, denoted as $(H_2O)_n$, were generated to be as spherical as possible.
Using double-$\zeta$ basis sets, both systems have a maximum of $d$-orbital angular momentum, while with triple-$\zeta$ basis sets, they have a maximum of $f$-orbital angular momentum.

In all tests, all screening thresholds were set to $10^{-10}$, apart from the opt-UM exchange threshold which was set to $10^{-5}$. It was verified that all implementations agreed on the final energy to within $10^{-5} E_h$.

\subsection{Single GPU performance comparison}

\begin{table*}
\resizebox{\textwidth}{!}{
\begin{tabular}{rlllllllllll}
\hline
\textbf{Software} & \textbf{Basis} & \textbf{gly$_{10}$} & \textbf{gly$_{20}$} & \textbf{gly$_{30}$} & \textbf{gly$_{40}$} & \textbf{gly$_{50}$} & \textbf{gly$_{60}$} & \textbf{gly$_{70}$} & \textbf{gly$_{80}$} & \textbf{gly$_{90}$} & \textbf{gly$_{100}$} \\ \hline
\multirow{3}{*}{\textbf{\texttt{TeraChem}}} & 6-31G* & 2.38 & 1.54 & 1.25 & 1.13 & 1.28 & 1.32 & 1.39 & 1.48 & 1.56 & 1.64 \\ \cline{2-12} 
 & 6-31G** & 2.29 & 1.33 & 1.18 & 1.25 & 1.33 & 1.45 & 1.57 & 1.65 & 1.74 & 1.83 \\ \cline{2-12} 
 & cc-pVDZ & 2.21 & 1.51 & 1.19 & 1.20 & 1.26 & 1.34 & 1.46 & 1.53 & 1.59 & 1.68 \\ \hline
\multirow{3}{*}{\textbf{\texttt{GPU4PySCF}}} & 6-31G* & 12.42 & 4.19 & 3.06 & 2.76 & 2.87 & 2.99 & 3.21 & 3.26 & 3.5 & 3.6 \\ \cline{2-12} 
 & 6-31G** & 7.72 & 3.71 & 2.76 & 2.93 & 3.21 & 3.35 & 3.64 & 3.74 & 3.94 & 4.16  \\ \cline{2-12} 
 & cc-pVDZ & 8.22 & 3.92 & 2.97 & 3.13 & 3.68 & 3.76 & 4.08 & 4.11 & 4.55 & 4.81 \\ \hline
\multirow{3}{*}{\textbf{\texttt{QUICK}}} & 6-31G* & 4.05 & 4.23 & 4.67 & 4.81 & 5.83 & 6.76 & 7.59 & 8.39 & - & - \\ \cline{2-12} 
 & 6-31G** & 3.93 & 4.20 & 4.57 & 5.83 & 6.99 & 8.08 & 9.09 & 9.97 & - & - \\ \cline{2-12} 
 & cc-pVDZ & 10.08 & 12.06 & 12.81 & 15.14 & 18.00 & - & - & - & - & - \\ \cline{2-12} 
\end{tabular}
\caption{Single A100 GPU Fock build speedup of the opt-Brc and opt-UM schemes over implementations in \texttt{TeraChem}, \texttt{QUICK} and \texttt{GPU4PySCF} on varying length polyglycine chains. The speedups are reported with respect to the minimum timing of the opt-Brc and opt-UM schemes, which is system and basis set dependent.}
\label{tab:gly speed ups} 
}\end{table*}

\begin{table*}
\resizebox{\textwidth}{!}{
\begin{tabular}{rlllllllllll}
\hline
\textbf{Software} & \textbf{Basis} & \textbf{(H$_2$O)$_{20}$} & \textbf{(H$_2$O)$_{40}$} & \textbf{(H$_2$O)$_{60}$} & \textbf{(H$_2$O)$_{80}$} & \textbf{(H$_2$O)$_{100}$} & \textbf{(H$_2$O)$_{120}$} & \textbf{(H$_2$O)$_{140}$} & \textbf{(H$_2$O)$_{160}$} & \textbf{(H$_2$O)$_{180}$} & \textbf{(H$_2$O)$_{200}$} \\ \hline
\multirow{3}{*}{\textbf{\texttt{TeraChem}}} & 6-31G* & 3.73 & 2.31 & 1.74 & 1.44 & 1.38 & 1.20 & 1.09 & 1.18 & 1.19 & 1.22  \\ \cline{2-12} 
 & 6-31G** & 2.94 & 1.92 & 1.53 & 1.18 & 1.05 & 1.07 & 1.05 & 1.16 & 1.16 & 1.21 \\ \cline{2-12} 
 & cc-pVDZ & 3.09 & 1.97 & 1.59 & 1.26 & 1.14 & 1.19 & 1.16 & 1.21 & 1.27 & 1.33 \\ \hline
\multirow{3}{*}{\textbf{\texttt{GPU4PySCF}}} & 6-31G* & 35.16 & 9.87 & 5.7 & 3.93 & 3.59 & 3.09 & 2.82 & 2.99 & 3.09 & 3.06 \\ \cline{2-12} 
 & 6-31G** & 23.55 & 6.76 & 4.24 & 3.21 & 2.91 & 2.8 & 2.84 & 3.04 & 3.16 & 3.27 \\ \cline{2-12} 
 & cc-pVDZ & 19.93 & 6.0 & 4.8 & 3.57 & 3.22 & 3.48 & 3.6 & 3.78 & 4.02 & 4.31 \\ \hline
\multirow{3}{*}{\textbf{\texttt{QUICK}}} & 6-31G* & 4.24 & 4.50 & 5.06 & 5.43 & 6.18 & 6.18 & 6.19 & 7.32 & 7.89 & 8.59 \\ \cline{2-12} 
 & 6-31G** & 4.19 & 4.57 & 5.33 & 5.36 & 5.53 & 6.31 & 6.88 & 8.06 & 8.69 & 9.42 \\ \cline{2-12} 
 & cc-pVDZ & 10.84 & 15.15 & 18.72 & 18.99 & 19.40 & 22.54 & - & - & - & - \\ \cline{2-12} 
\end{tabular}
\caption{Single A100 GPU Fock build speedup of the opt-Brc and opt-UM schemes over implementations in \texttt{TeraChem}, \texttt{QUICK} and \texttt{GPU4PySCF} on varying size water clusters. The speedups are reported with respect to the minimum timing of the opt-Brc and opt-UM schemes, which is system and basis set dependent.}\label{tab:water speed ups} 
}\end{table*}

To evaluate the relative efficiency of the optimized implementations developed in this work, the single GPU performance was benchmarked against existing GPU accelerated Fock builds in \texttt{TeraChem} (v1.96H)~\cite{ufimtsev_quantum_2008, ufimtsev_quantum_2009-1, ufimtsev_quantum_2009}, \texttt{QUICK} (v2.0)~\cite{manathunga_quick-2403_2024, manathunga_quantum_2023, miao_acceleration_2015, manathunga_harnessing_2021} and \texttt{GPU4PySCF} (v0.7.7)~\cite{li_introducing_2024, sun_recent_2020}. The \texttt{TeraChem} implementation incorporates the original UM09 scheme, as discussed earlier in this Article. The \texttt{QUICK} implementation uses an HGP integral scheme but also draws inspiration from the UM09 scheme with presorting and thread walking for effective screening~\cite{miao_acceleration_2015}. \texttt{GPU4PySCF} implements the Rys Quadrature~\cite{rys_computation_1983, dupuis_evaluation_1976} ERI evaluation scheme and uses the Brc21 scheme for integral class creation and screening. We note that \texttt{GPU4PySCF} v1.0 was recently released however we observed a significant performance improvement using v0.7.7.
Both \texttt{GPU4PySCF} and \texttt{TeraChem} use a differential Fock build to improve the screening of later SCF iterations. \texttt{TeraChem} also supports a mixed precision implementation. Both of these features were disabled in the following results to provide a meaningful comparison between schemes. These features will be addressed in future work.

Figure \ref{fig:comparisons} shows single-iteration Fock build timings for our opt-Brc and opt-UM schemes, as well as for \texttt{QUICK}, \texttt{TeraChem} and \texttt{GPU4PySCF} for increasing size polyglicines and water clusters, across the 6-31G*, 6-31G**, cc-pVDZ and cc-pVTZ basis sets.  These single-iteration timings were obtained by averaging over 10 SCF iterations and are presented on a log scale to make the small and large molecular scale behaviour visible.

Figure \ref{fig:comparisons} demonstrates that our opt-UM scheme is the fastest implementation for large molecular systems while our opt-Brc scheme is the fastest implementation for smaller systems. 
For double-$\zeta$ basis sets, the performance crossover point between the two schemes occurs at approximately 3{,}000 basis functions for both linear (polyglycine) and globular (water cluster) systems.

For all systems and basis sets tested, our optimised implementation with the minimum execution time presents a speed up against the implementations in both \texttt{QUICK}, \texttt{TeraChem} and \texttt{GPU4PySCF}. These speedups are presented in Tables \ref{tab:gly speed ups} and \ref{tab:water speed ups}, for the polyglycine and water systems, respectively.

For the polyglycine systems using the cc-pVDZ basis set, we observed average speedups of $1.5\times$, $4.3\times$, and $13.6\times$ compared to \texttt{TeraChem}, \texttt{GPU4PySCF}, and \texttt{QUICK}, respectively. Similarly, for the water systems using the cc-pVDZ basis set, the average speedups were $1.5\times$, $5.7\times$, and $17.6\times$ over \texttt{TeraChem}, \texttt{GPU4PySCF}, and \texttt{QUICK}, respectively.

Considering both the polyglycine and water systems with the cc-pVDZ basis set, the minimum speedups observed were $1.14\times$, $2.97\times$, and $10.08\times$ with respect to \texttt{TeraChem}, \texttt{GPU4PySCF}, and \texttt{QUICK}, respectively. The maximum speedups for the same systems and basis set were $3.09\times$, $19.93\times$, and $22.54\times$ relative to \texttt{TeraChem}, \texttt{GPU4PySCF}, and \texttt{QUICK}, respectively.
 
As shown in Fig.~\ref{fig:comparisons}, panels \ref{fig:gly-cc-pVTZ} and \ref{fig:water-cc-pVTZ}, the novel $f$ function capability was benchmarked using the cc-pVTZ basis set. Since \texttt{TeraChem} does not yet support $f$ functions, only \texttt{QUICK} and \texttt{GPU4PySCF} were used for comparison. However, \texttt{QUICK} failed to converge for glycine chains longer than gly\textsubscript{5}. These tests were conducted on 80GB A100 GPUs to accommodate the significant register spills associated with the $f$ function implementation. 
We note a significant improvement in the relative performance of \texttt{GPU4PySCF} with the triple-$\zeta$ basis set, while still being outperformed by our opt-Brc scheme. This can be attributed to the Rys Quadrature ERI scheme which is known to yield good performance on high angular momentum integral classes~\cite{gill_molecular_1994}.
Notably, the opt-Brc scheme significantly outperforms the opt-UM scheme beyond the previously observed $3{,}000$ basis function crossover point.
The superior performance of the opt-Brc scheme can be partially attributed to the increased number of basis functions per atom, which reduces sparsity and decreases the effectiveness of the screening implemented in the opt-UM scheme. Furthermore, beyond requiring the computation of complex integrals involving $f$-type CGFs, usage of the cc-pVTZ basis set requires the evaluation of a significantly larger number of medium- to high-contraction degree integrals with lower angular momentum CGFs. For these integrals, the HGP scheme within the opt-Brc implementation requires fewer FLOPs than the MMD scheme. This can result in a significant performance advantage of opt-Brc over opt-UM, especially for integrals involving highly contracted functions, which, as demonstrated later in Fig. \ref{fig:TZ_int_class_comp}, are identified as the main computational bottleneck. 

As shown in Fig.~\ref{fig:comparisons}, with increasing system size, the relative performance of the opt-UM scheme improves over the opt-Brc one.

\subsubsection{Comparison to \texttt{LibintX}}
\begin{table}[b]
\begin{tabular}{lcc}
\textbf{Molecule} & \texttt{LibintX} & opt-UM \\ \hline
Taxol & 4 & 2.75 \\ \hline
Olestra & 10.1 & 4.8 \\ \hline
gly\textsubscript{120} & 36.3 & 6.41 \\ \hline
Crambin & 95.7 & 64.2 \\ \hline
Ubiquitin & 298.1 & 81.2
\end{tabular}
\caption{Single-iteration $\bm{K}$ formation timings (in seconds) for \texttt{LibintX} and our opt-UM scheme on a selection of molecules using the 6-31G* basis set. The tests were run on a single V100 GPU.}\label{tab:libintx-comparison}
\end{table}

\texttt{LibintX}~\cite{asadchev_3-center_2024, asadchev_high-performance_2023} is a GPU accelerated quantum chemistry library which uses the MMD scheme for ERI evaluation. The algorithm in \texttt{LibintX} leverages the MMD formulation to compute ERIs via matrix-matrix multiplications, at the cost of reduced utilization of $E_{\bm{p}}^{\bm{a}\bm{b}}$ sparsity, although it does exploit the sparsity from Eq.~\eqref{eq:MMD-observation}.

\citeauthor{asadchev_3-center_2024} report the $\bm{K}$ matrix formation timings per SCF iteration on a single V100 GPU~\cite{asadchev_3-center_2024}. Table \ref{tab:libintx-comparison} presents a comparison of these timings with those of our $\bm{K}$ algorithm within the opt-UM scheme implementation.

Due to memory constraints on V100 GPUs (see Section \ref{sec:f functions}), we limit our comparison to the 6-31G* basis set. It is important to note that the \texttt{LibintX} results were generated using a purely spherical basis set, which employs fewer basis functions than our Cartesian basis. We note that \citeauthor{asadchev_3-center_2024} report that the implementation in \texttt{LibintX} was not fully optimized; however, we present these comparisons for completeness.

As shown in Table \ref{tab:libintx-comparison}, our implementation achieves an average speedup of 2.9$\times$. A maximum speedup of 5.7$\times$ was observed for the linear gly\textsubscript{120} system, indicating that linear $\bm{K}$ algorithm is particularly advantageous in such cases.

\subsection{Strong Scaling}

\begin{figure}[t]
    \centering
    \includegraphics[width=1\linewidth]{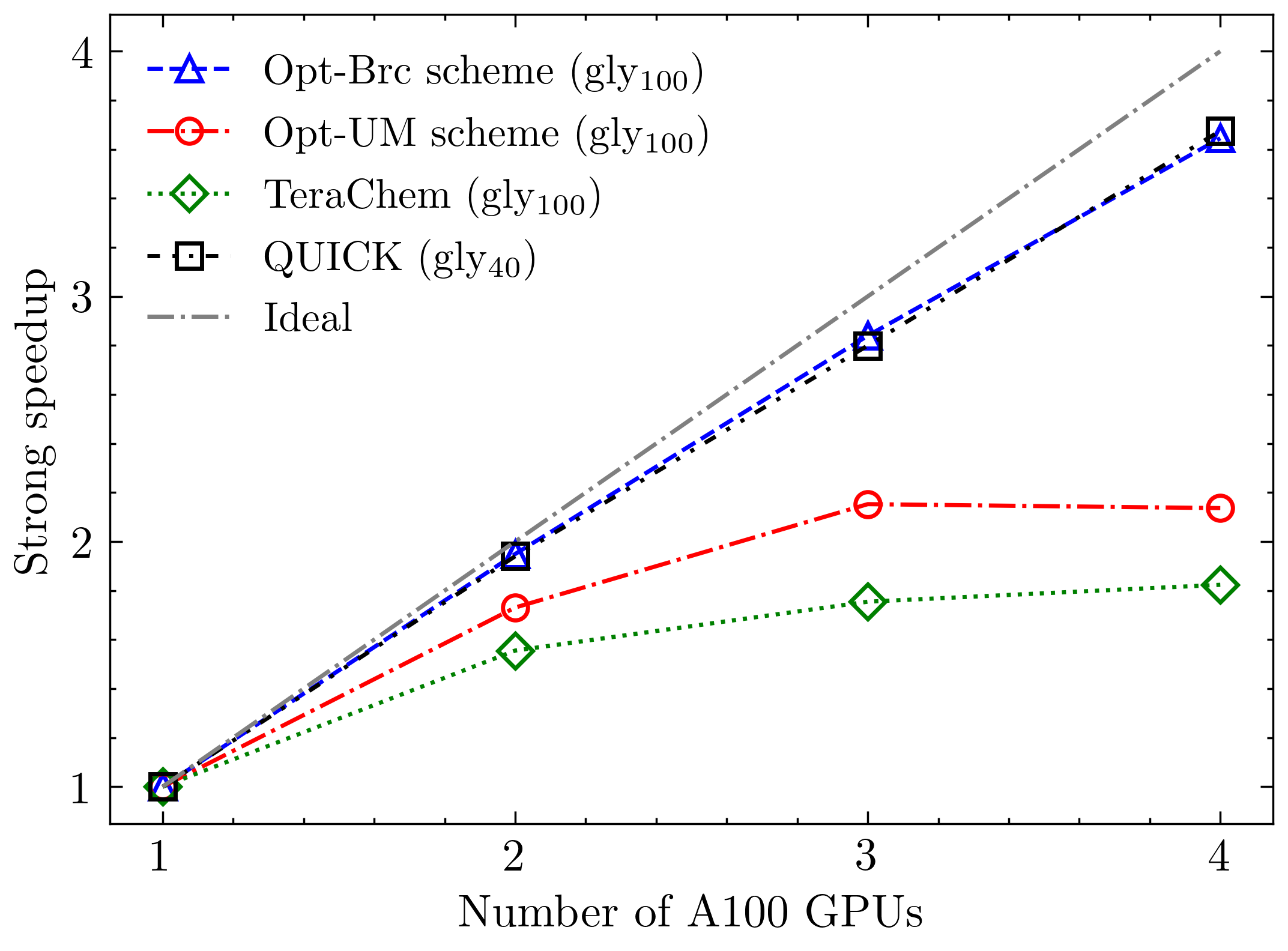}
    \caption{Strong scaling performance of Fock build timings of opt-Brc, opt-UM, \texttt{TeraChem} and \texttt{QUICK} implementations on 1 to 4 A100 GPUs. Tests were conducted on glycine chains with the cc-pVDZ basis set.}
    \label{fig:strong_scaling}
\end{figure}

To analyze the parallel efficiency of the four implementations in \texttt{QUICK}, \texttt{TeraChem}, and our opt-Brc and opt-UM schemes, we measured the overall strong speedup per SCF iteration using an increasing number of GPUs.

Polyglycine chains with the cc-pVDZ basis were selected and executed on up to four A100 GPUs for this strong scaling analysis. The multi-GPU implementation in \texttt{QUICK} encountered a segmentation fault on polyglycine chains longer than gly\textsubscript{50}; hence, \texttt{QUICK} strong scaling results were obtained with gly\textsubscript{40}. In contrast, \texttt{TeraChem} and our opt-Brc and opt-UM implementations were run on gly\textsubscript{100} to ensure the system was sufficiently large for effective parallelization.  Figure \ref{fig:strong_scaling} presents the strong scaling results.

We observed excellent strong speedup for both implementations based on the HGP scheme, namely the \texttt{QUICK} and the opt-Brc algorithms, with both achieving over 91\% total parallel efficiency on four GPUs. In contrast, the \texttt{Terachem} and opt-UM algorithms, which are both based on the  MMD scheme, exhibited significantly poorer strong scaling, achieving only approximately 50\% parallel efficiency on four GPUs. 

This discrepancy of parallel efficiency can be attributed to the additional overheads inherent in the MMD-based algorithms. These overheads, which are not distributed among the GPUs and are performed at each iteration, include the sorting of primitive shell-pair batches prior to the $\bm{J}$ computation and the calculation of the upper bound for the density matrix elements required for the linear $\bm{K}$ algorithm. There is potential for further parallelization of these components in future work. Additionally, these overheads are likely to be further mitigated in larger systems, where a greater portion of execution time is spent evaluating $\bm{J}$ and $\bm{K}$.

\subsection{Single Node comparison to CPU software}
\begin{figure}[t]
    \centering
    \includegraphics[width=1\linewidth]{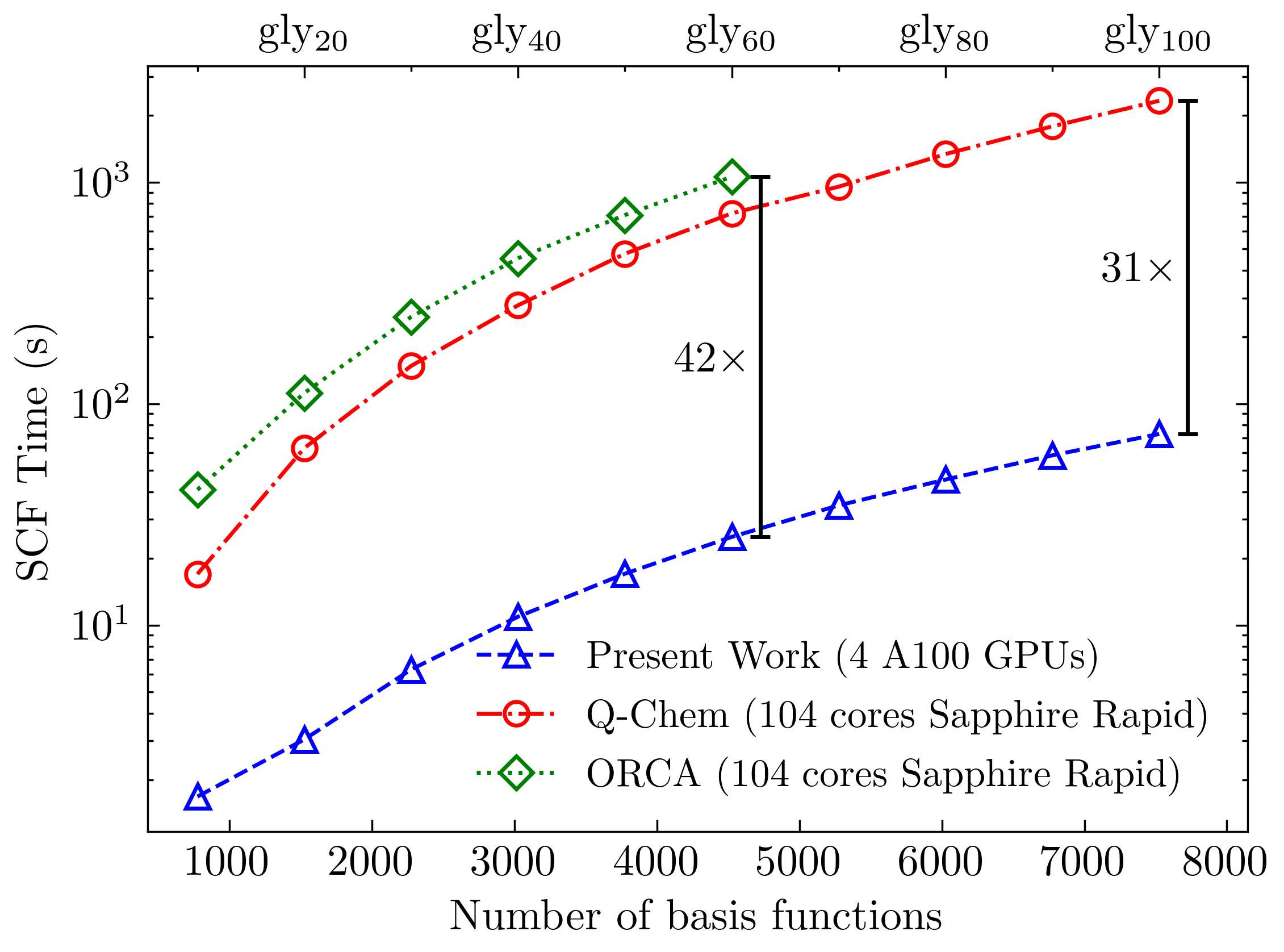}
    \caption{Single node performance comparison of total SCF computation time of the opt-Brc scheme with CPU packages \texttt{Q-Chem} and \texttt{ORCA} on varying length glycine chains with the cc-pVDZ basis set.}
    \label{fig:cpu_comparison}
\end{figure}

In this subsection, we compare the performance of our opt-Brc and opt-UM Fock build algorithms with those of the SCF procedures implemented in the widely-used CPU packages \texttt{ORCA} (v5.0.4)~\cite{neese_orca_2012} and \texttt{Q-Chem} 6.0~\cite{shao_advances_2015}. 

\texttt{ORCA} and \texttt{Q-Chem} were benchmarked on 2$\times$52 core Intel Xeon Platinum 8470Q (Sapphire Rapid) 2.1 GHz CPUs with and 512 GB DDR4 RAM. Our opt-Brc and opt-UM schemes were run on a single node containing 4$\times$A100 GPUs. For our implementation we report the minimum total SCF time between the opt-Brc and opt-UM algorithm.

The timing comparisons are shown in Fig. \ref{fig:cpu_comparison}.
We observe large speedups of our GPU code over both CPU implementations, with a 42$\times$ speedup over \texttt{ORCA} on gly\textsubscript{60} and a 31$\times$ speed up over \texttt{Q-Chem} on gly\textsubscript{100}. Note that \texttt{ORCA} was only benchmarked to gly\textsubscript{60} as it ran out of memory on gly\textsubscript{70}. These speedups are expected to further improve for longer polyglycine chains due to the favorable screening observed with our implementation.

The Sapphire Rapids CPUs have a Thermal Design Power (TDP) of approximately 350 W, whereas each A100 GPU has a TDP of 400 W. Consequently, a node with 4 GPUs consumes 2.3 times more power than a node with 2 CPUs. Thus, the speedups of 42$\times$ over \texttt{ORCA} and 31$\times$ over \texttt{Q-Chem} correspond to improvements in power efficiency of approximately $18\times$ and $12\times$, respectively.

\subsection{Additional Benchmarks}

\subsubsection{J and K scaling}
\begin{figure}[t]
    \centering
    \includegraphics[width=1\linewidth]{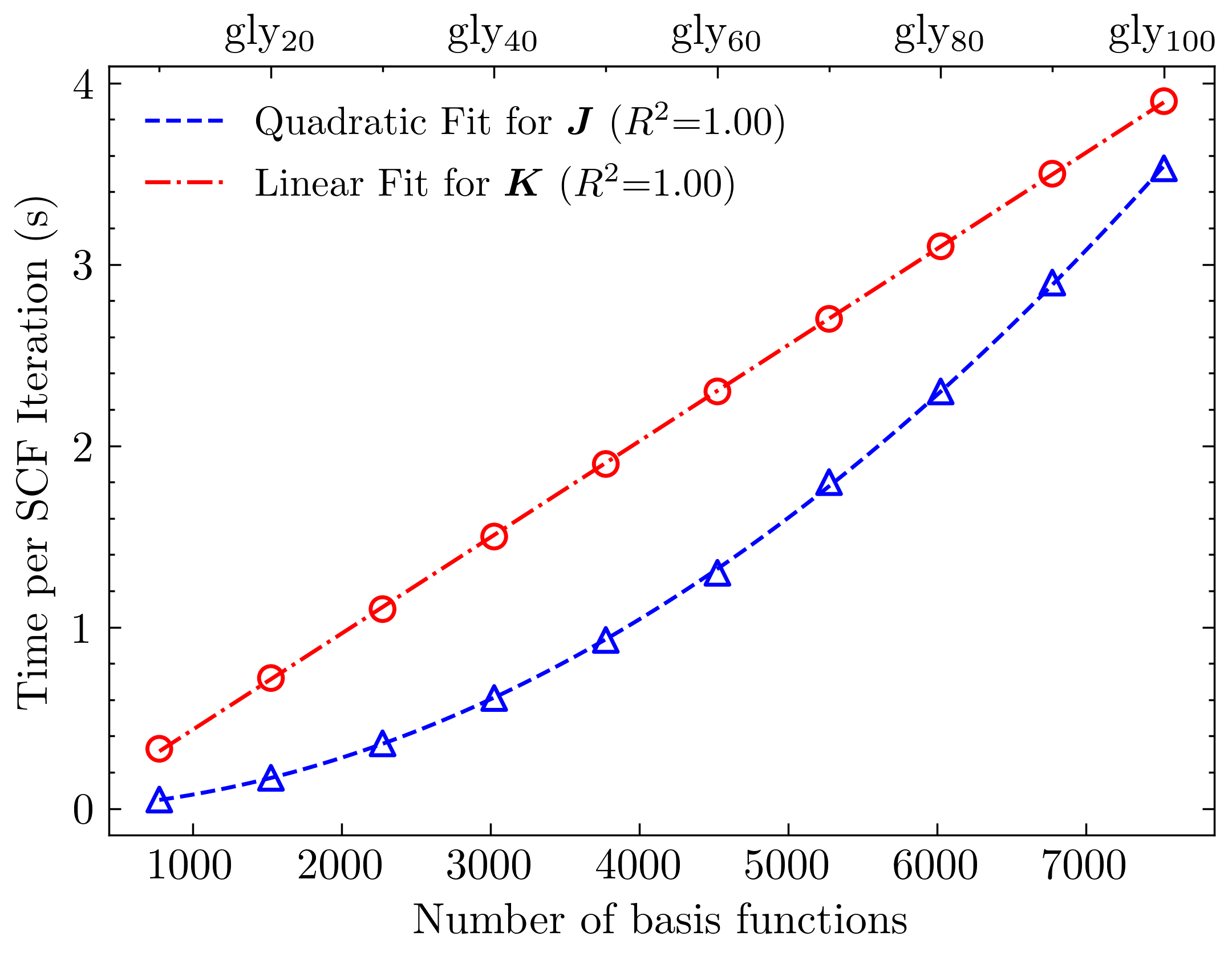}
    \caption{Scaling of the Coulomb ($\bm{J}$) and exchange repulsion ($\bm{K}$) matrix formation timings within the opt-UM scheme for varying length polyglycine chains. The basis set adopted was cc-pVDZ. }
    \label{fig:J_K_scaling}
\end{figure}

The primary objective of exploiting the exponential decay of the density matrix within the opt-UM scheme was to achieve linearly scaling $\bm{K}$ matrix formation time with system size.

Additionally, we anticipated the $\bm{J}$ matrix computation to scale quadratically with system size due to the ERI screening. 

To verify this, we measured the single-iteration timings for both the Coulomb and exchange matrices across polyglycine chains of increasing lengths using the cc-pVDZ basis set. As shown in Fig. \ref{fig:J_K_scaling}, the results confirm highly accurate linear and quadratic scaling for the exchange ($\bm{K}$) and Coulomb ($\bm{J}$) matrix computations, respectively.

\subsubsection{Integral Class Timings}
\begin{figure}[h!]
    \centering
    \begin{subfigure}[b]{\linewidth}
        \centering
        \includegraphics[width=1\linewidth]{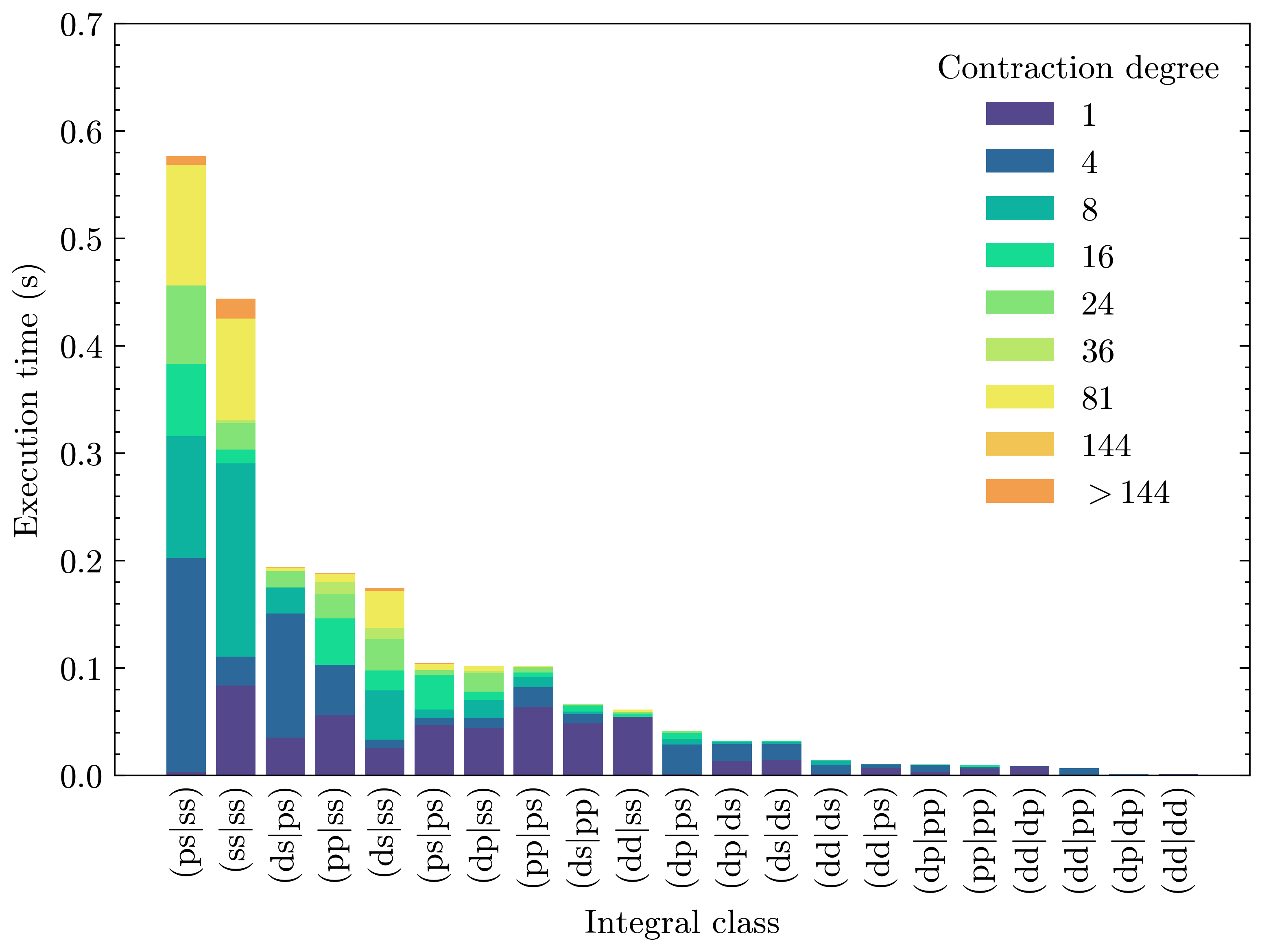}
        \caption{$(H_2O)_{100}$ with the cc-pVDZ basis set.}
        \label{fig:DZ_int_class_comp}
    \end{subfigure}    
    \begin{subfigure}[b]{\linewidth}
        \centering
        \includegraphics[width=1\linewidth]{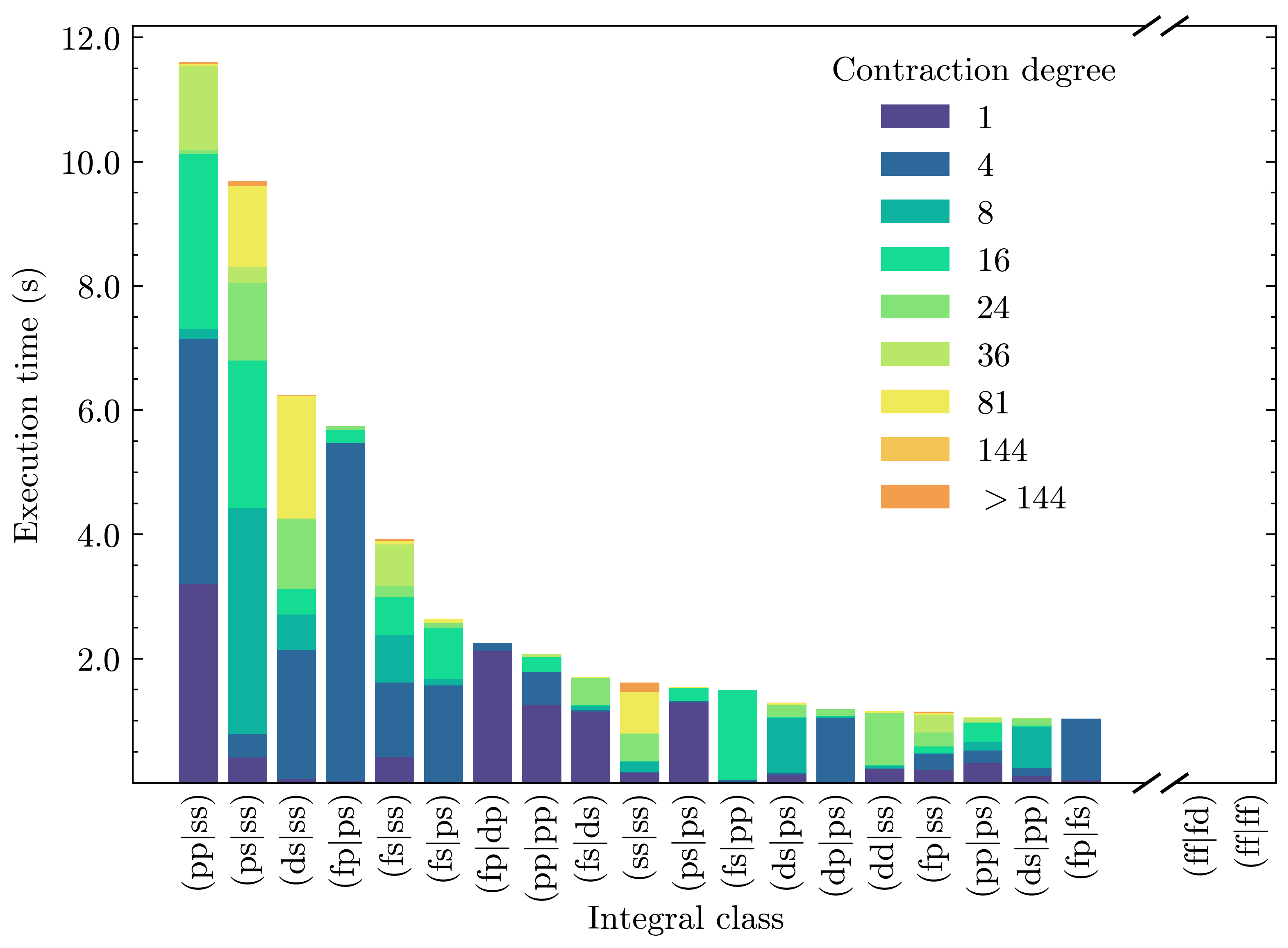}
        \caption{$(H_2O)_{100}$ with the cc-pVTZ basis set.}
        \label{fig:TZ_int_class_comp}
    \end{subfigure}
    
    \caption{Execution time per integral class for the second SCF iteration of $(H_2O)_{100}$ with double- and triple-$\zeta$ basis sets. Results were generated with the opt-Brc scheme on a single stream of an A100 GPU.}
    \label{fig:int_class_comp}
\end{figure}
In this subsection, we analyse how the time required to compute different contracted integrals classes contributes to the total Fock build runtime.

The opt-Brc scheme was selected to generate these results so the influence of contraction degree could be compared. For meaningful results, all kernels were launched on a single stream. This decreases the performance of the implementation due to the loss of multi-stream parallelism but ensures all kernel launches have access to the same GPU resources. As detailed in Section \ref{sec:barca-opt}, we assign shell pairs to integral classes based on the number of \textit{significant} primitive pairs rather than the product of contraction degrees. Without this optimisation, we would observe a significantly larger proportion of execution time being required for highly contracted integral classes.

Figures \ref{fig:DZ_int_class_comp} and \ref{fig:TZ_int_class_comp} present the execution time per integral class for a single SCF iteration of $(H_2O)_{100}$ when using the the cc-pVDZ and cc-pVTZ basis sets, respectively. 
We note that as the opt-Brc scheme fully exploits the 8-way ERI symmetry, the symmetric integral classes have a comparatively lower execution time.
We observe that despite the complexity of evaluating high angular momentum integral classes, the sheer number of lower angular momentum shells results in the associated integral classes to clearly dominate execution time. For example, the the 6 symmetry-unique integral classes that contain only $s$ and $p$ angular momentum take up over 60\% of the total execution time with the cc-pVDZ basis set.

\subsubsection{SCF timings}
\begin{table}
\centering
\resizebox{\columnwidth}{!}{%
\begin{tabular}{llcclc}
Molecule & Basis set & \begin{tabular}[c]{@{}c@{}}No. Basis \\ functions\end{tabular} & \begin{tabular}[c]{@{}c@{}}No. \\ GPUs\end{tabular} & Scheme & \begin{tabular}[c]{@{}c@{}}Time per SCF \\ Iteration (s)\end{tabular} \\ \hline
\multirow{2}{*}{Taxol} & \multirow{2}{*}{cc-pVTZ} & \multirow{2}{*}{2970} & 1 & opt-Brc & 36.5 \\ \cline{4-6} 
 &  &  & 4 & opt-Brc & 9.2 \\ \hline
\multirow{2}{*}{Olestra} & \multirow{2}{*}{cc-pVDZ} & \multirow{2}{*}{4015} & 1 & opt-Brc & 7.6 \\ \cline{4-6} 
 &  &  & 4 & opt-Brc & 2.0 \\ \hline
\multirow{2}{*}{gly120} & \multirow{2}{*}{cc-pVDZ} & \multirow{2}{*}{9025} & 1 & opt-UM & 11.2 \\ \cline{4-6} 
 &  &  & 4 & opt-UM & 5.7 \\ \hline
\multirow{2}{*}{Crambin} & \multirow{2}{*}{cc-pVDZ} & \multirow{2}{*}{6390} & 1 & opt-UM & 55.4 \\ \cline{4-6} 
 &  &  & 4 & opt-Brc & 18.9 \\ \hline
\multirow{2}{*}{Ubiquitin} & \multirow{2}{*}{6-31G*} & \multirow{2}{*}{10273} & 1 & opt-UM & 62.9 \\ \cline{4-6} 
 &  &  & 4 & opt-UM & 24.7
\end{tabular}
}
\caption{Average of 10 SCF iteration timings for standard benchmark molecules on a variety of basis sets on 1 and 4 A100 GPUs.}
\label{tab:benchmarks}
\end{table}

We have provided per SCF iteration timings for a variety of standard molecules for completeness. The opt-Brc and opt-UM schemes were both run on each input and the minimum execution time is reported in Table \ref{tab:benchmarks}. Due to the near ideal strong scaling of the opt-Brc scheme it is sometimes preferable to use the opt-Brc scheme with multi GPU execution. In contrast, the opt-UM scheme excels for large molecular systems when executed on a single GPU.

%====================
\section{Conclusions}
%====================
This Article presented substantial advancements in the construction of the Fock matrix using GPUs, introducing two optimized algorithms: opt-Brc and opt-UM. Both algorithms have been fine-tuned to enhance integral screening, exploit sparsity and symmetry, and extend the capabilities for HF computations up to $f$-type angular momentum functions.

Specifically, the integral screening techniques incorporated into both algorithms significantly reduce computational costs. Efficient Cauchy-Schwarz screening and the exploitation of the nearsightedness of the density matrix for exchange matrix calculations are central to these improvements. The opt-Brc algorithm fully utilizes the eight-fold permutational symmetry of ERIs, while the opt-UM algorithm leverages the separation of Coulomb and exchange contributions to enhance screening and performance. A notable achievement is the opt-UM algorithm's linear scaling exchange matrix assembly, achieved by exploiting the exponential decay of the density matrix, resulting in significant performance improvement over the previous UM09 scheme. 

The algorithms also extend the capabilities for HF calculations to include $f$-type angular momentum functions, enabling more accurate and diverse HF calculations. A dedicated code generator handles the complex tree search problem of generating recurrence relations for high angular momentum classes, ensuring near-optimal evaluation strategies and reducing manual coding errors.

Performance benchmarks on NVIDIA A100 GPUs demonstrate that our implementations in EXESS outperform the current leading GPU and CPU Fock build implementations found in \texttt{TeraChem}, \texttt{QUICK}, \texttt{GPU4PySCF}, \texttt{LibIntX}, \texttt{ORCA}, and \texttt{Q-Chem}. 

For linear polyglycine chains using the cc-pVDZ basis set, our algorithms achieve average speedups of 1.5$\times$, 4.3$\times$, and 13.6$\times$ over \texttt{TeraChem}, \texttt{GPU4PySCF}, and \texttt{QUICK}, respectively. Similarly, for globular water clusters, the speedups are 1.5$\times$, 5.7$\times$, and 17.6$\times$. The maximum speedups observed for cc-pVDZ are 3.09$\times$, 19.93$\times$, and 22.54$\times$ relative to \texttt{TeraChem}, \texttt{GPU4PySCF}, and \texttt{QUICK}, respectively. For the 6-31G* and 6-31G** basis sets, even larger maximum speedups were observed with respect to \texttt{TeraChem}, \texttt{GPU4PySCF}, and \texttt{QUICK}, across the polyglycine and water test molecular systems.

Using the 6-31G* basis set, the $\bm{K}$ assembly algorithm in the opt-UM scheme was also benchmarked against the $\bm{K}$ matrix implementation in \texttt{LibIntX}. In this case, the opt-UM exchange scheme achieved average and maximum speedups of 2.9$\times$ and 5.7$\times$ over \texttt{LibIntX}.

Strong scaling analysis reveal over 91\% parallel efficiency on four GPUs for opt-Brc, highlighting its typically superior performance for multi-GPU execution.

When compared to widely-used CPU-based software such as \texttt{ORCA} and \texttt{Q-Chem}, our algorithms offer speedups of up to 42$\times$ and 31$\times$, respectively, translating to power efficiency improvements of up to 18.3$\times$.

Future enhancements currently under development include mixed precision implementations, differential Fock build, and additional algorithms for low and high-angular momentum integrals. Development of high-performance multi-GPU molecular gradients is also underway. 

%==========================
\section{Acknowledgements}
%==========================
The authors thank the National Energy Research Scientific Computing Center (NERSC), a Department of Energy Office of Science User Facility using award ERCAP0026496 for resource allocation on the Perlmutter supercomputer.
EP and RS acknowledge the National Industry PhD program, the Department of Education and QDX technologies for providing additional funding.
The authors also thank the NCMAS and ANUMAS computational allocation schemes for access to the Gadi supercomputer at NCI and the Setonix supercomputer at the Pawsey Supercomputing Centre.

\bibliography{extra_refs, references}

\end{document}